\documentclass[a4paper,twocolumn,11pt,accepted=2024-03-22]{quantumarticle}
\pdfoutput=1
\usepackage[utf8]{inputenc}
\usepackage[english]{babel}
\usepackage[T1]{fontenc}
\usepackage{tikz}

\usepackage{graphicx}
\usepackage{color}
\usepackage{braket}
\usepackage{verbatim}
\usepackage{amsmath}
\usepackage{amsfonts}
\usepackage{mathtools}
\usepackage{braket}
\usepackage{float}
\usepackage{hyperref}
\usepackage{lipsum}
\usepackage{dsfont}
\usepackage{tabularx,ragged2e}

\DeclarePairedDelimiter\abs{\lvert}{\rvert}

\begin{document}

\title{Loss-tolerant architecture for quantum computing with quantum emitters}
\author{Matthias C. L\"{o}bl}
\affiliation{Center for Hybrid Quantum Networks (Hy-Q), The Niels Bohr Institute, University of Copenhagen, Blegdamsvej 17, DK-2100 Copenhagen {\O}, Denmark}
\email{matthias.loebl@nbi.ku.dk}
\orcid{0000-0003-4744-3119}

\author{Stefano Paesani}
\affiliation{Center for Hybrid Quantum Networks (Hy-Q), The Niels Bohr Institute, University of Copenhagen, Blegdamsvej 17, DK-2100 Copenhagen {\O}, Denmark}
\affiliation{NNF Quantum Computing Programme, Niels Bohr Institute, University of Copenhagen, Denmark.}
\orcid{0000-0001-5709-0906}
\email{stefano.paesani@nbi.ku.dk}

\author{Anders S. S\o{}rensen}
\affiliation{Center for Hybrid Quantum Networks (Hy-Q), The Niels Bohr Institute, University of Copenhagen, Blegdamsvej 17, DK-2100 Copenhagen {\O}, Denmark}
\orcid{0000-0003-1337-9163}
\email{anders.sorensen@nbi.ku.dk}

\begin{abstract}
We develop an architecture for measurement-based quantum computing using photonic quantum emitters. The architecture exploits spin-photon entanglement as resource states and standard Bell measurements of photons for fusing them into a large spin-qubit cluster state. The scheme is tailored to emitters with limited memory capabilities since it only uses an initial non-adaptive (ballistic) fusion process to construct a fully percolated graph state of multiple emitters. By exploring various geometrical constructions for fusing entangled photons from deterministic emitters, we improve the photon loss tolerance significantly compared to similar all-photonic schemes.

\end{abstract}

\maketitle

\section{Introduction}
Measurement-based quantum computing requires the generation of large graph states (cluster states) followed by measurements on them~\cite{Raussendorf2001, Raussendorf2003, Briegel2009}. This approach is particularly promising for photonic systems since quantum operations can be implemented with linear optics and photon detectors only. The required photonic cluster states can be created from small resource states by connecting them through so-called fusion processes~\cite{Kieling2007,GimenoSegovia2015,Pant2019}. This approach, however, has several challenges: first, fusions are probabilistic and consume photonic qubits~\cite{Knill2001}. Furthermore, photons travel at immense speed, which necessitates long delay lines to implement conditional feedback operations~\cite{Bombin2021}. Most critically, however, photons are easily lost which puts stringent bounds on the required photon efficiency. Recent schemes require efficiencies $>97\%$~\cite{GimenoSegovia2015,Bartolucci2021} which is above the typical values of photonic platforms~\cite{Zhong2018, paesani2020, Uppu2020, Tomm2021}. Furthermore, schemes using few-photon resource states require boosting~\cite{Grice2011,Ewert2014} the fusion success probability with ancillary photons~\cite{GimenoSegovia2015, Pant2019, Bartolucci2021}, which further complicates the experimental implementation. Improved architectures have been developed based on larger initial resource states~\cite{Bartolucci2021}, but even few-photon states can only be made with a low probability using the typically employed parametric down conversions sources~\cite{Walther2005,Zhong2018, paesani2020}. New approaches are thus required to make the generation of large-scale cluster states experimentally feasible.

\begin{figure}[t]
\includegraphics[width=1.0\columnwidth]{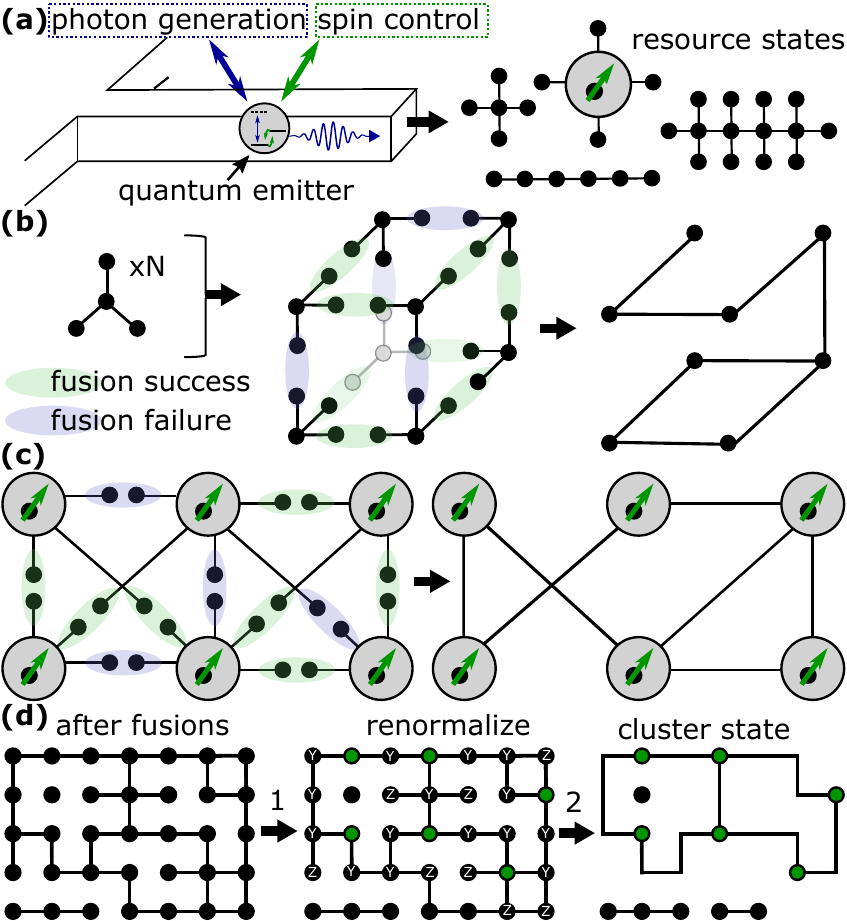}
\caption{\label{fig:scheme}\textbf{(a)} A quantum emitter in a waveguide can used to generate different types of resource states~\cite{Lindner2009,Tiurev2021}. \textbf{(b)} Several star-shaped photonic resource states are arranged on a simple cubic lattice. Fusions of the resource states are performed by using the photons on the {\it leaves} of the resource states. Sufficiently many successful fusions generate a large connected cluster state. \textbf{(c)} Several star-shaped resource states with a quantum emitter spin as the central qubit can be fused into a distributed spin graph state. \textbf{(d)} As the fusions succeed with a finite probability, parts of the desired graph state are missing. The state can be renormalized into a cluster state with a well-defined lattice (here a square lattice) via Y and Z measurements~\cite{Herr2018}.}
\end{figure}

Fortunately, there are promising new methods to generate large resource states~\cite{Gheri1998,Buterakos2017}. In particular quantum emitters, such as quantum dots, even enable generating them in a deterministic and thus scalable way~\cite{Lindner2009,Schwartz2016,Tiurev2021,Coste2022,Cogan2023}. At the same time, these emitters have high photon efficiencies on-chip~\cite{Acari2014,Scarpelli2019} and end-to-end~\cite{Uppu2020, Tomm2021}. Indeed, the largest photonic resource states that have ever been generated are GHZ-states made with a quantum emitter~\cite{Thomas2022}. Several resource states that can be generated with a single emitter are shown in Fig.~\ref{fig:scheme}(a). Fig.~\ref{fig:scheme}(b) illustrates a lattice on which photonic resource states are geometrically arranged and fused. Besides purely photonic resource states, quantum emitters can also generate states where a stationary spin is entangled with the photons~\cite{Delteil2016,Stockill2017,Appel2022}. Here, we exploit these properties to construct an architecture, which uses star-shaped resource states (locally equivalent to GHZ states) with a central spin qubit. The spin is entangled with several photonic {\it leaf} qubits where the connections represent the entanglement properties (see Fig.~\ref{fig:scheme}(c)). From these resource states, we generate a large spin cluster state via rotated type-II fusions (Bell measurements on the photons)~\cite{Browne2005}. We propose an implementation with semiconductor quantum dots~\cite{Warburton2013,Lodahl2015}, but the scheme can also be applied to atoms~\cite{Thomas2022} or color centers~\cite{Bernien2013}. By using spin qubits as the building blocks of the cluster state, we remove the need to implement feedback on flying qubits as well as unheralded loss of the qubits in the final cluster state. The latter poses a challenge to purely photonic approaches~\cite{Morley2017}. The proposed hybrid approach combines the advantages of spin-based and photon-based platforms: spins in quantum dots are excellent photon emitters with coherence times~\cite{Zaporski2022,Nguyen2023} much longer than qubit initialization, readout, and manipulation~\cite{Xu2007,Antoniadis2022,Press2008}, but so far no clear strategy existed for how to scale these systems. In our proposal, the photons provide a fast link between the static spins enabling full-scale quantum computing.

Previous proposals for generating large spin-spin entanglement use a repeat-until-success strategy~\cite{Barrett2005,Lim2005,Duan2005} that creates an immense overhead in the number of qubits~\cite{Barrett2005} and requires long coherence times of the qubits. This overhead can be circumvented to some extent with more than one qubit per emitter, one for storage and one for generating the entanglement~\cite{Choi2019,Denning2019,Pompili2021}, but this introduces additional experimental complexity. Our cluster-state generation does not use repeat-until-success and all fusions are performed in one shot (ballistically), enabling a high overall clock speed.
In contrast to previously proposed ballistic schemes~\cite{GimenoSegovia2015,Pant2019}, our architecture can operate loss-tolerantly without boosting~\cite{Grice2011,Ewert2014} the fusion success probability with ancillary photons. It only requires rotated type-II fusions~\cite{Browne2005,Gimeno2016} where success, failure, and photon loss are heralded on the detection pattern. All these features keep the experimental overhead low and make our approach particularly feasible.

To optimize the tolerance to photon loss, we explore several lattices on which star-shaped resource states are arranged and fused. Since there are no locality constraints for entanglement generated by Bell measurements on photons, we consider lattices in several dimensions and search for ideal lattices with a discrete optimization algorithm.

\section{Building lattices by fusing resource states}
Our approach starts by arranging resource states on a \textit{fusion lattice} as illustrated in Fig.~\ref{fig:scheme}(c). Here the resource state is represented as a star-shaped graph with the central spin in the middle and photon as the {\it leaves} of the star. It is defined by:
\begin{equation}
\prod_{j=1}^N C_{sj}\left(\ket{+}_s\otimes\ket{+}^{\otimes N}\right),
\end{equation}
where the first qubit is a spin ($s$) and the $N$ other qubits are photons. The state $\ket{+}=\frac{1}{\sqrt{2}}(\ket{0}+\ket{1})$ is the $+1$ eigenstate of the Pauli X operator~\footnote{We use the following definition of the Pauli-matrices: $X=\big(\begin{smallmatrix}0 & 1\\1 & 0\end{smallmatrix}\big)$, $Y=\big(\begin{smallmatrix}0 & -i\\i & 0\end{smallmatrix}\big)$, $Z=\big(\begin{smallmatrix}1 & 0\\0 & -1\end{smallmatrix}\big)$}. $C_{sj}$ represents a controlled-$Z$ gate between the spin and photon number $j$ (see Appendix~\ref{appendix:resource} for details). The central spin qubit of the star-shaped state will be part of the final cluster state and the photons on the leaves are used to perform fusions with other resource states. The fusion that we consider succeeds with a probability of $p_s=0.5$~\cite{Browne2005,Gimeno2016} in which case a connection between two central qubits is established~\cite{GimenoSegovia2015} (see Appendix~\ref{appendix:fusion} for the required setup, known as rotated type-II fusion). When the fusion fails, there is no connection and the leaves are erased. When enough fusions succeed, a large connected entangled state is created, which is a resource for measurement-based quantum computing~\cite{Raussendorf2001,Kieling2007}. The required success probability of the fusions can be quantified by a so-called percolation threshold~\cite{Kieling2007,GimenoSegovia2015,Pant2019}: when the success probability of fusion is below the bond percolation threshold of the lattice, the generated graph state consists of many small pieces and is useless for quantum computing (that applies for instance to the two-dimensional honeycomb lattice~\cite{Sykes1964}). Above the percolation threshold, a cluster state with a large connected component spanning the entire lattice is generated. Such a graph state can then be renormalized into a lattice~\cite{Herr2018} by local Pauli measurements~\cite{Hein2004,Hein2006} as illustrated in Fig.~\ref{fig:scheme}(d). From here quantum computation can be performed by measurements on the generated graph~\cite{Raussendorf2001}.

\begin{figure*}[t]
\includegraphics[width=1.0\textwidth]{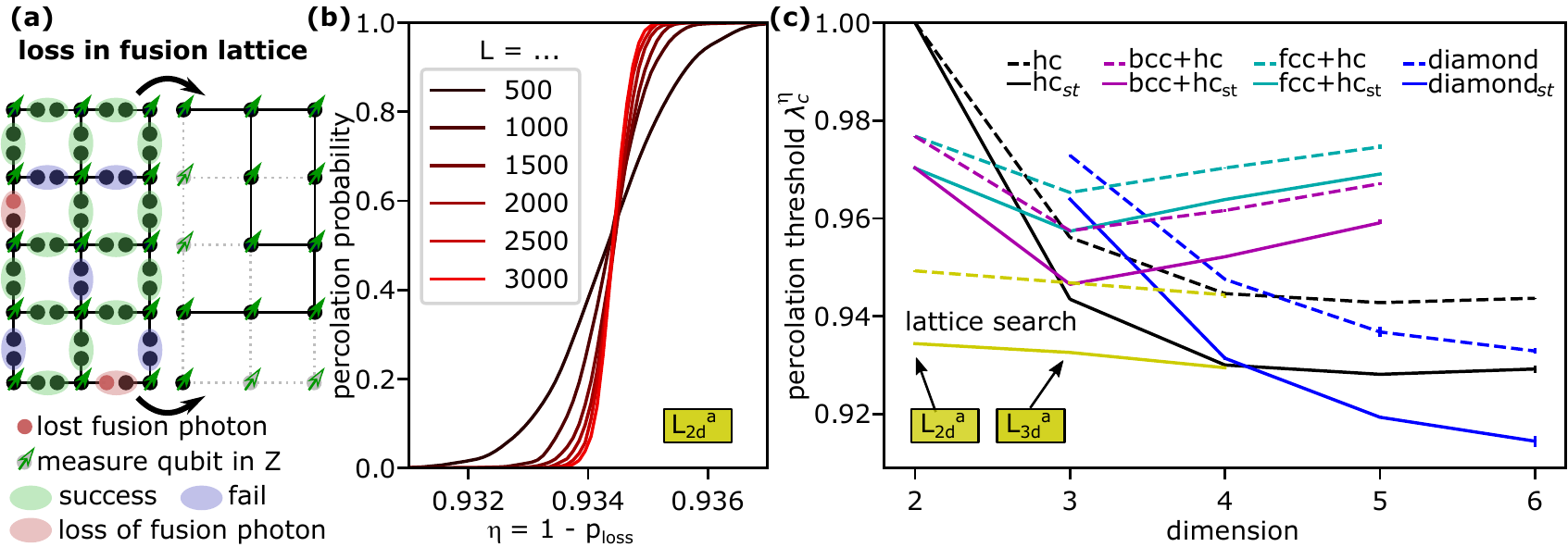}
\caption{\label{fig:2}\textbf{(a)} Loss on a fusion lattice where star-shaped resource states are geometrically arranged and fused. When a fusion photon is lost, the neighborhoods of both fusion photons are removed from the graph state by Z-measurements to retain a pure quantum state. \textbf{(b)} Percolation simulation for estimating the tolerance to photon loss. The simulated lattice is two-dimensional and has been optimized for a low percolation threshold. Every curve is an average of over $10^3$ simulations. Above a percolation threshold $\lambda_c^{\eta}$, the probability for a connection between the edges of a $d$-dimensional lattice of size $L^d$ approaches unity as $L\rightarrow \infty$. \textbf{(c)} Percolation thresholds $\lambda_c^{\eta}$ (minimum required value for $\eta=1-p_{loss}$). The thresholds are obtained by extrapolating the simulation results of lattices with finite sizes $L$ (see part (b)) towards infinity using the method from Ref.~\cite{VanderMarck1998}. The considered lattices are hypercubic (hc), diamond, body-centered cubic combined with hc (bcc+hc), face-centered cubic combined with hc (fcc+hc), as well as lattices obtained by an optimization ($L_{2d}^a$, $L_{3d}^a$). The simulations are plotted for all-photonic star-shaped resource states (dashed lines) as well as resource states with a quantum emitter spin as a static (st) central qubit (solid lines).}
\end{figure*}

However, photon losses make it challenging to generate a large-scale percolated cluster state as they lead to a mixed quantum state~\cite{Hein2006}. To retain a pure state, the neighborhood of every lost qubit is removed from the graph state by measurements in the $Z$-basis~\cite{GimenoSegovia2015}. When fusion photons are lost, the fusion heralds the loss but not which of the two fusion qubits was lost. Therefore, the neighborhoods of both fusion qubits are removed from the graph state~\cite{GimenoSegovia2015} (see Fig.~\ref{fig:2}(a)).

In our model, every edge of the fusion lattice has two fusion photons and a loss thus occurs with probability $1 - (1-p_{loss})^2$ per edge, where $p_{loss}$ is the probability that an individual photon is lost ($p_{loss}$ is assumed to be the same for all photons). In contrast, fusion success occurs with probability $(1-p_{loss})^2\cdot p_s$ and fusion failure with $(1-p_{loss})^2\cdot (1-p_s)$ (both without loss). With this model, we compute percolation thresholds for photon loss: only when the photon efficiency $\eta=1-p_{loss}$ is above the percolation threshold $\lambda_c^{\eta}$, a percolated graph state with a large connected component is created.

Our simulations generally consist of three steps: we first build a lattice defining the used resource states and the performed fusions, we then apply the described model for fusion failure and photon loss, and finally we evaluate if the resulting graph state percolates the lattice (if it is a spanning cluster).

\section{Lattice construction}
The loss tolerance of the constructions described above depends on which lattice the resource states are geometrically arranged and fused. Therefore, we consider various fusion lattices and analyze their tolerance to photon loss. Our lattice construction is based on the $d$-dimensional hypercubic lattice $\mathbb{Z}^d$~\cite{VanderMarck1998, Kurzawski2012}. Every point of the lattice represents the central qubit of a star-shaped resource state and the edges represent which fusions are performed. Fig.~\ref{fig:scheme}(b) corresponds to a simple cubic fusion lattice for instance. We represent all the connections to the neighbors of a lattice point (qubit) by integer vectors $\vec{z}\in\mathbb{Z}^d$ that illustrate the corresponding geometric differences. We consider graphs with a local neighborhood such that for every connection vector $\vec{z}$, its maximum integer value per dimension is restricted. Initially, the maximum integer value per dimension is set to $k=1$ ($\vec{z} \in \Set{0,\pm1}^d \backslash \Set{0}$). Lattices of this type are for instance the hypercubic or the fcc lattice. Further lattices such as the $d$-dimensional brickwork representation of the diamond lattice~\cite{GimenoSegovia2015,VanderMarck1998} can be obtained by removing particular edges from these lattices. A more detailed description of all these lattices can be found in Appendix~\ref{appendix:lattice_def}. To validate our implementation of these lattices, we compare the corresponding classical site- and bond-percolation thresholds~\cite{VanderMarck1998, Kurzawski2012} in Appendix~\ref{appendix:classical}.

Note that, in practice, the construction of these high-dimensional lattices is obtained in a standard set-up by collapsing them in the (3+1)-dimensional space. Such a construction is suitable for the photonic platform as the long-range links that are generated in such a collapse can be implemented with minimal loss via optical fibers.

\section{Simulating photon loss} As a metric for the loss tolerance of a graph state construction, we use the percolation probability (a cluster spanning from one edge of the simulated lattice to the other) when fusion failures and photon losses are probabilistically applied (see Fig.~\ref{fig:2}(a)). A corresponding percolation simulation of a two-dimensional lattice is shown in Fig.~\ref{fig:2}(b). In all simulations, we consider the emitter to be the central qubit of the star-shaped resource state (see Fig.~\ref{fig:scheme}(c)). This type of qubit cannot be lost, in contrast to the fused photonic qubits. The percolation thresholds for several multi-dimensional lattices are shown in Fig.~\ref{fig:2}(c), and the corresponding values can be found in Appendix~\ref{appendix:thresh}. For the best lattices, the percolation thresholds $\lambda_c^{\eta}$ are below $0.94$ showing that a photon loss probability $p_{loss}$ of about $1-\lambda_c^{\eta}>6\%$ can be tolerated.\footnote{A detailed description of the algorithm that we use to efficiently perform the percolation simulations for photon loss is provided in Ref.~\cite{Lobl2023b}.} Associated numerical results for the size of the largest connected cluster component are given in Appendix~\ref{appendix:size}.

For completeness, we also simulate the loss tolerance when using purely photonic resource states where the central qubits can also suffer unheralded loss~\cite{Morley2017}, reported as dashed lines in Fig.~\ref{fig:2}(c). Note that, in contrast to our spin-based approach, the obtained loss thresholds for the all-photonic cases only provide necessary conditions for creating a useful cluster state after lattice renormalization~\cite{Kieling2007,Herr2018}.

 For the considered lattices, Fig.~\ref{fig:2}(c) shows that the hypercubic lattices perform best in dimensions 3 and 4 whereas the diamond lattices are ideal for higher dimensions. A remarkable feature is the dependence of the percolation thresholds on the dimension. For the hypercubic (hc) lattices and the lattices fcc+hc, bcc+hc, we observe an optimum dimension where the corresponding fusion lattice has the lowest percolation threshold. For the diamond lattice, it is not obvious where the optimum is but we expect such an optimum: with increasing dimension, the vertex degree of the corresponding lattices increases. On the one hand, a higher vertex degree is desirable because more fusion attempts lead to a higher chance of establishing connections in the cluster state. On the other hand, in the presence of photon loss, a too high vertex degree (more fusion photons) is problematic since the loss of a fusion qubit makes the central qubit useless (it must be measured in the Z-basis, see Fig.~\ref{fig:2}(a)). We further analyze this point in Appendix~\ref{appendix:size} in the context of the largest connected cluster state component. A dimension where the percolation threshold of a certain lattice type reaches a minimum is a feature of the fragile nature of entanglement. This differs from classical bond- and site-percolation where the percolation thresholds always decrease when adding more bonds to the same lattice.

\section{Lattice optimization}
Fig.~\ref{fig:2}(c) illustrates that different lattices show quite different performances under photon loss. Therefore, we search for ideal lattices in a given dimension by a discrete optimization algorithm. As before, we virtually place vertices on a hypercubic lattice $\mathbb{Z}^d$ and represent the connections by a set $E$ of integer vectors: $E\subseteq \Set{\vec{z}\in\mathbb{Z}^d: \abs{z_i}\leq k} \setminus \Set{0}$, where $k\in\mathbb{N}^+$ bounds the maximum number of steps that a connection vector $\vec{z}$ makes per dimension. The lattice representation by connection vectors allows constructing lattices with so-called complex neighborhoods~\cite{Malarz2005,Xun2020b} yet it is more general and flexible than the typical approach of specifying lattices by nearest neighbor connections~\cite{Malarz2005,Xun2020b}. Therefore, the chosen representation is particularly suited for optimizations within a large space of geometries. We only consider lattices where all nodes have an identical neighborhood, so the presence of a vector $\vec{z}\in E$ implies that there is also a connection to a neighbor in the opposite direction $-\vec{z}\in E$. Therefore, all lattices that we consider have an even vertex degree.

\begin{figure*}
\includegraphics[width=2.0\columnwidth]{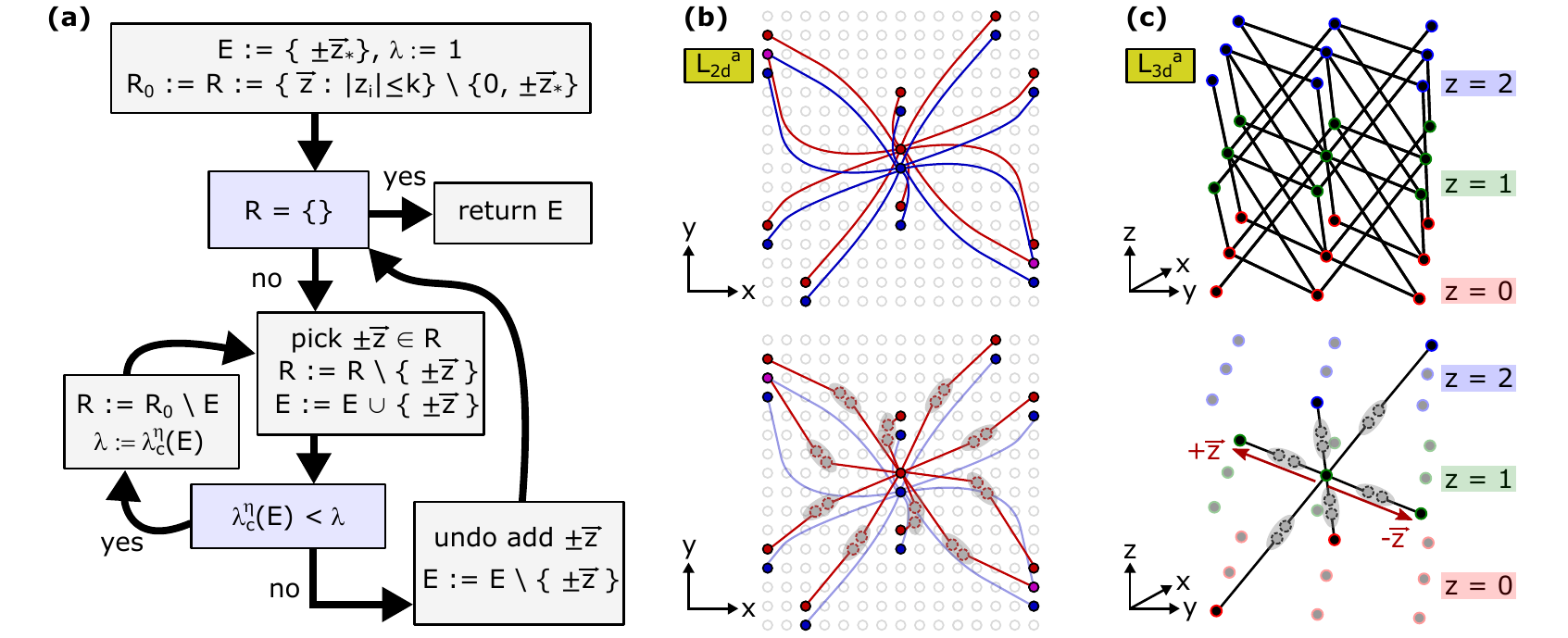}
\caption{\label{fig:lattices}\textbf{(a)} Flow chart of the algorithm used to optimize the loss tolerance of fusion lattices. The lattice is represented by a set of nodes on $\mathbb{Z}^d$ together with connection vectors between the nodes specified by the set $E$. The algorithm adds pairs of connection vectors $\pm\vec{z}\in R$ to $E$ trying to improve the percolation threshold $\lambda_c\left(E\right)$. \textbf{(b)} Two-dimensional fusion lattice $L_{2d}^a$ with high loss tolerance. In the abstract representation of the fusion lattice (upper part), every connection is a fusion, and every node is the center of a resource state. For one exemplary star-shaped resource state, an explicit representation of the fusions is shown (lower part). The red and blue graph structures represent the fusions connected to two qubits in the center of the plot. Every node on the two-dimensional square lattice (gray) has the same connection pattern. \textbf{(c)} Optimized fusion lattice $L_{3d}^a$ in three dimensions. Every qubit (node) is linked to eight neighbors by fusions (edges). An exemplary pair of connection vectors $\pm\vec{z}$ is illustrated in the lower part.}
\end{figure*}

A flow chart of our algorithm is shown in Fig.~\ref{fig:lattices}(a). The algorithm starts from a lattice where every node only has two connection vectors $\pm\vec{z}_*$ ($E=\{\vec{z}_*, -\vec{z}_*\}$) that are randomly chosen from the set $\Set{\vec{z}\in\mathbb{Z}^d: \abs{z_i}\leq k} \setminus \Set{0}$. Next, two random vectors $\pm\vec{z}$ are taken from a reservoir of vectors $R\equiv\Set{\vec{z}\in\mathbb{Z}^d: \abs{z_i}\leq k} \setminus \Set{0, \pm \vec{z}_*}$ and added to the lattice (resp. moved from $R$ to $E$). If the percolation threshold $\lambda_c^{\eta}\left(E\right)$ improves by adding these vectors, the algorithm adds further vectors from $R$. In contrast, if the percolation threshold gets worse, the algorithm removes the two most recently added vectors from $E$ and adds another pair of vectors $\pm\vec{z}\in R$ instead. If adding any vector from $R$ just makes the percolation threshold worse, the algorithm terminates. In this case, $E$ likely represents a good lattice that cannot be improved by adding any other vector remaining in $R$~\footnote{Note, that one could instead start from a lattice with all possible vectors and remove vectors randomly. However, when starting from a lattice with a very high vertex degree, the algorithm often does not succeed. The reason is that the percolation threshold can only be determined with limited accuracy which can cause the algorithm to terminate too early.}.

In two and three dimensions, we find lattices that show better performance in comparison to the simple lattices studied above. These lattices ($L_{2d}^a$ and $L_{3d}^a$) are shown in Fig.~\ref{fig:lattices}(b,c) and the vectors representing the different lattices are given in Appendix~\ref{appendix:lattice}. They can already tolerate up to 6.5\% photon loss when the standard type-II fusions only succeed with $p_s=0.5$. The lattice $L_{2d}^a$ has an interesting spider structure with some local connection plus some long legs making connections to nodes that are further away. We believe that this lattice has a high loss tolerance because it imitates properties of higher dimensional lattices. In a high-dimensional lattice, a loss may break a connection along a certain dimension, but connections in the other dimensions find a way around it. The long connections of the spider-lattice may play a similar role by bridging local interruptions caused by photon loss. In four dimensions and with a photon as the central qubit, the algorithm finds a lattice with a performance that is almost identical to the 4d hypercubic lattice. Remarkably, when the central qubit is a spin, the algorithm finds exactly the 4d hypercubic lattice (resp. an identical lattice, up to distortions).

\section{Discussion}
We have proposed a scheme for generating large spin cluster states for measurement-based quantum computing. The basic building blocks are star-shaped resource states that can be generated on demand by various types of quantum emitters with a spin~\cite{Lindner2009, Tiurev2021}, a resource that is experimentally in reach~\cite{Schwartz2016, Thomas2022, Cogan2023, Coste2022}. We fuse these resource states performing all fusions simultaneously (ballistically) allowing a fast clock rate for generating large-scale entanglement. We only use non-boosted rotated type-II fusion~\cite{Browne2005}, and find fusion lattices in two and higher dimensions that can tolerate about $6-8\%$ photon loss. This loss tolerance is an improvement by more than a factor of two compared to similar ballistic schemes~\cite{GimenoSegovia2015} that are all-photonic and an order of magnitude compared to intrinsically fault-tolerant schemes~\cite{Bartolucci2021,Paesani2022} even though those schemes assume boosted fusion and some even employ larger resource states. The reason for the improvement is a combination of our lattice optimization and the fact that some adaptiveness is moved to the required lattice renormalization step~\cite{Kieling2007, Herr2018}. Alternatively, repeat-until-success schemes can be used to further improve the loss-tolerance by a divide-and-conquer strategy~\cite{Barrett2005, Duan2005} which, however, creates a significant overhead since large parts of the lattice construction may have to be repeated in case of fusion failure. Furthermore, the need to do operations sequentially puts a high demand on quantum memories. Another proposal~\cite{Choi2019} applies a repeat-until-success scheme to generate entanglement between one type of quantum emitter spins which is swapped to a second type of spins upon success. Such a highly adaptive scheme can improve the loss tolerance to arbitrary values, but increases the requirements on the emitters, since they now need to contain two spins. These schemes can therefore only be applied to certain emitters, such as NV-centers~\cite{Pompili2021}, but are e.g. not directly applicable to quantum dots. A suitable compromise might be a minimally adaptive repeat-until-success scheme where each entanglement attempt can be repeated if it is unsuccessful. Allowing a fixed number $n_{r}$ of fusion attempts would decrease the overall rate by the constant factor $n_{r}$, but would increase the fusion success probability to $1-1/2^{n_{r}}$ in the absence of losses (see Ref.~\cite{Lobl2023b}). In contrast to such approaches, the scheme discussed in this paper optimizes the loss tolerance while at the same time keeping the requirement on the photon emitters to a minimum. 

Our results can guide experimental work towards scalable quantum computing with quantum emitters and provide performance thresholds that such emitters need to meet. To further reduce the experimental requirements we see several possible extensions of our work: (1) We only consider a certain type of lattices and more general classes of lattices could be simulated by using combinatorial tiling theory~\cite{Newman2020} or quasi-periodic tilings~\cite{Kramer1989}. (2) In our approach, lattices are built from star-shaped (GHZ) states. Using more complex resource states might further improve the loss tolerance and potentially also enable the exploitation of loss and error-tolerant sub-spaces~\cite{Bell2022}. However, generating such states is more involved and might require direct spin-spin gates between emitters~\cite{Economou2010,Michaels2021,Li2022}. (3) Our scheme uses quantum emitter spins as the qubits and thus will require a large number of emitters. It would be interesting to estimate this number and investigate methods to reduce it by recycling qubits, keeping only part of the cluster state active during computation~\cite{Morley2017} (4) Here we focus on loss tolerance. Integrating the approach with techniques for quantum error correction against logical errors can yield a fully fault-tolerant architecture. In this regard, it is encouraging that fault tolerance is related to similar percolation concepts~\cite{Stace2009,Auger2018,Hastings2014} facilitating its integration with the current approach. Furthermore, error correction benefits from high-dimensional structures, which also favors the developed architecture~\cite{Terhal2015, Breuckmann2016, Breuckmann2021}.

\section{Ackowledgements}
We are grateful for discussions with Love A. Pettersson, Yu-Xiang Zhang, and Peter Lodahl. We are grateful for financial support from Danmarks Grundforskningsfond (DNRF 139, Hy-Q Center for Hybrid Quantum Networks). S.P. acknowledges funding from the Cisco University Research Program Fund (nr. 2021-234494), from the Marie Skłodowska-Curie Fellowship project QSun (nr. 101063763), from the VILLUM FONDEN research grant VIL50326, and support from the NNF Quantum Computing Programme.

\bibliography{lit}

\begin{thebibliography}{10}

\bibitem{Raussendorf2001}
Robert Raussendorf and Hans~J. Briegel.
\newblock ``A one-way quantum computer''.
\newblock \href{https://dx.doi.org/10.1103/PhysRevLett.86.5188}{Phys. Rev.
  Lett. {\bf 86}, 5188--5191}~(2001).

\bibitem{Raussendorf2003}
Robert Raussendorf, Daniel~E. Browne, and Hans~J. Briegel.
\newblock ``Measurement-based quantum computation on cluster states''.
\newblock \href{https://dx.doi.org/10.1103/PhysRevA.68.022312}{Phys. Rev. A
  {\bf 68}, 022312}~(2003).

\bibitem{Briegel2009}
Hans~J Briegel, David~E Browne, Wolfgang D{\"u}r, Robert Raussendorf, and
  Maarten Van~den Nest.
\newblock ``Measurement-based quantum computation''.
\newblock \href{https://dx.doi.org/10.1038/nphys1157}{Nat. Phys. {\bf 5},
  19--26}~(2009).

\bibitem{Kieling2007}
K.~Kieling, T.~Rudolph, and J.~Eisert.
\newblock ``Percolation, renormalization, and quantum computing with
  nondeterministic gates''.
\newblock \href{https://dx.doi.org/10.1103/PhysRevLett.99.130501}{Phys. Rev.
  Lett. {\bf 99}, 130501}~(2007).

\bibitem{GimenoSegovia2015}
Mercedes Gimeno-Segovia, Pete Shadbolt, Dan~E. Browne, and Terry Rudolph.
\newblock ``From three-photon {G}reenberger-{H}orne-{Z}eilinger states to
  ballistic universal quantum computation''.
\newblock \href{https://dx.doi.org/10.1103/PhysRevLett.115.020502}{Phys. Rev.
  Lett. {\bf 115}, 020502}~(2015).

\bibitem{Pant2019}
Mihir Pant, Don Towsley, Dirk Englund, and Saikat Guha.
\newblock ``Percolation thresholds for photonic quantum computing''.
\newblock \href{https://dx.doi.org/10.1038/s41467-019-08948-x}{Nat. Commun.
  {\bf 10}, 1070}~(2019).

\bibitem{Knill2001}
Emanuel Knill, Raymond Laflamme, and Gerald~J Milburn.
\newblock ``A scheme for efficient quantum computation with linear optics''.
\newblock \href{https://dx.doi.org/10.1038/35051009}{Nature {\bf 409},
  46--52}~(2001).

\bibitem{Bombin2021}
Hector Bombin, Isaac~H Kim, Daniel Litinski, Naomi Nickerson, Mihir Pant,
  Fernando Pastawski, Sam Roberts, and Terry Rudolph.
\newblock ``Interleaving: Modular architectures for fault-tolerant photonic
  quantum computing''~(2021).
\newblock
  url:~\href{https://doi.org/10.48550/arXiv.2103.08612}{doi.org/10.48550/arXiv.2103.08612}.

\bibitem{Bartolucci2021}
Sara Bartolucci, Patrick Birchall, Hector Bombin, Hugo Cable, Chris Dawson,
  Mercedes Gimeno-Segovia, Eric Johnston, Konrad Kieling, Naomi Nickerson,
  Mihir Pant, et~al.
\newblock ``Fusion-based quantum computation''.
\newblock \href{https://dx.doi.org/10.1038/s41467-023-36493-1}{Nat. Commun.
  {\bf 14}, 912}~(2023).

\bibitem{Zhong2018}
Han-Sen Zhong, Yuan Li, Wei Li, Li-Chao Peng, Zu-En Su, Yi~Hu, Yu-Ming He, Xing
  Ding, Weijun Zhang, Hao Li, Lu~Zhang, Zhen Wang, Lixing You, Xi-Lin Wang,
  Xiao Jiang, Li~Li, Yu-Ao Chen, Nai-Le Liu, Chao-Yang Lu, and Jian-Wei Pan.
\newblock ``12-photon entanglement and scalable scattershot boson sampling with
  optimal entangled-photon pairs from parametric down-conversion''.
\newblock \href{https://dx.doi.org/10.1103/PhysRevLett.121.250505}{Phys. Rev.
  Lett. {\bf 121}, 250505}~(2018).

\bibitem{paesani2020}
S.~Paesani, M.~Borghi, S.~Signorini, A.~Ma\"inos, L.~Pavesi, and A.~Laing.
\newblock ``Near-ideal spontaneous photon sources in silicon quantum
  photonics''.
\newblock \href{https://dx.doi.org/10.1038/s41467-020-16187-8}{Nat. Commun.
  {\bf 11}, 2505}~(2020).

\bibitem{Uppu2020}
Ravitej Uppu, Freja~T Pedersen, Ying Wang, Cecilie~T Olesen, Camille Papon,
  Xiaoyan Zhou, Leonardo Midolo, Sven Scholz, Andreas~D Wieck, Arne Ludwig,
  et~al.
\newblock ``Scalable integrated single-photon source''.
\newblock \href{https://dx.doi.org/10.1126/sciadv.abc8268}{Sci. Adv. {\bf 6},
  eabc8268}~(2020).

\bibitem{Tomm2021}
Natasha Tomm, Alisa Javadi, Nadia~Olympia Antoniadis, Daniel Najer,
  Matthias~Christian L{\"o}bl, Alexander~Rolf Korsch, R{\"u}diger Schott,
  Sascha~Ren{\'e} Valentin, Andreas~Dirk Wieck, Arne Ludwig, et~al.
\newblock ``A bright and fast source of coherent single photons''.
\newblock \href{https://dx.doi.org/10.1038/s41565-020-00831-x}{Nat.
  Nanotechnol. {\bf 16}, 399--403}~(2021).

\bibitem{Grice2011}
W.~P. Grice.
\newblock ``Arbitrarily complete bell-state measurement using only linear
  optical elements''.
\newblock \href{https://dx.doi.org/10.1103/PhysRevA.84.042331}{Phys. Rev. A
  {\bf 84}, 042331}~(2011).

\bibitem{Ewert2014}
Fabian Ewert and Peter van Loock.
\newblock ``$3/4$-efficient bell measurement with passive linear optics and
  unentangled ancillae''.
\newblock \href{https://dx.doi.org/10.1103/PhysRevLett.113.140403}{Phys. Rev.
  Lett. {\bf 113}, 140403}~(2014).

\bibitem{Walther2005}
Philip Walther, Kevin~J Resch, Terry Rudolph, Emmanuel Schenck, Harald
  Weinfurter, Vlatko Vedral, Markus Aspelmeyer, and Anton Zeilinger.
\newblock ``Experimental one-way quantum computing''.
\newblock \href{https://dx.doi.org/10.1038/nature03347}{Nature {\bf 434},
  169--176}~(2005).

\bibitem{Gheri1998}
K.~M. Gheri, C.~Saavedra, P.~T\"orm\"a, J.~I. Cirac, and P.~Zoller.
\newblock ``Entanglement engineering of one-photon wave packets using a
  single-atom source''.
\newblock \href{https://dx.doi.org/10.1103/PhysRevA.58.R2627}{Phys. Rev. A {\bf
  58}, R2627--R2630}~(1998).

\bibitem{Buterakos2017}
Donovan Buterakos, Edwin Barnes, and Sophia~E. Economou.
\newblock ``Deterministic generation of all-photonic quantum repeaters from
  solid-state emitters''.
\newblock \href{https://dx.doi.org/10.1103/PhysRevX.7.041023}{Phys. Rev. X {\bf
  7}, 041023}~(2017).

\bibitem{Lindner2009}
Netanel~H. Lindner and Terry Rudolph.
\newblock ``Proposal for pulsed on-demand sources of photonic cluster state
  strings''.
\newblock \href{https://dx.doi.org/10.1103/PhysRevLett.103.113602}{Phys. Rev.
  Lett. {\bf 103}, 113602}~(2009).

\bibitem{Schwartz2016}
Ido Schwartz, Dan Cogan, Emma~R Schmidgall, Yaroslav Don, Liron Gantz, Oded
  Kenneth, Netanel~H Lindner, and David Gershoni.
\newblock ``Deterministic generation of a cluster state of entangled photons''.
\newblock \href{https://dx.doi.org/10.1126/science.aah4758}{Science {\bf 354},
  434--437}~(2016).

\bibitem{Tiurev2021}
Konstantin Tiurev, Pol~Llopart Mirambell, Mikkel~Bloch Lauritzen,
  Martin~Hayhurst Appel, Alexey Tiranov, Peter Lodahl, and Anders~S\o{}ndberg
  S\o{}rensen.
\newblock ``Fidelity of time-bin-entangled multiphoton states from a quantum
  emitter''.
\newblock \href{https://dx.doi.org/10.1103/PhysRevA.104.052604}{Phys. Rev. A
  {\bf 104}, 052604}~(2021).

\bibitem{Coste2022}
N.~Coste, D.A. Fioretto, N.~Belabas, S.C. Wein, P.~Hilaire, R.~Frantzeskakis,
  M.~Gundin, B.~Goes, N.~Somaschi, M.~Morassi, et~al.
\newblock ``High-rate entanglement between a semiconductor spin and
  indistinguishable photons''.
\newblock \href{https://dx.doi.org/10.1038/s41566-023-01186-0}{Nature Photonics
  {\bf 17}, 582--587}~(2023).

\bibitem{Cogan2023}
Dan Cogan, Zu-En Su, Oded Kenneth, and David Gershoni.
\newblock ``Deterministic generation of indistinguishable photons in a cluster
  state''.
\newblock \href{https://dx.doi.org/10.1038/s41566-022-01152-2}{Nat. Photon.
  {\bf 17}, 324--329}~(2023).

\bibitem{Acari2014}
M.~Arcari, I.~S\"ollner, A.~Javadi, S.~Lindskov~Hansen, S.~Mahmoodian, J.~Liu,
  H.~Thyrrestrup, E.~H. Lee, J.~D. Song, S.~Stobbe, and P.~Lodahl.
\newblock ``Near-unity coupling efficiency of a quantum emitter to a photonic
  crystal waveguide''.
\newblock \href{https://dx.doi.org/10.1103/PhysRevLett.113.093603}{Phys. Rev.
  Lett. {\bf 113}, 093603}~(2014).

\bibitem{Scarpelli2019}
L.~Scarpelli, B.~Lang, F.~Masia, D.~M. Beggs, E.~A. Muljarov, A.~B. Young,
  R.~Oulton, M.~Kamp, S.~H\"ofling, C.~Schneider, and W.~Langbein.
\newblock ``99\% beta factor and directional coupling of quantum dots to fast
  light in photonic crystal waveguides determined by spectral imaging''.
\newblock \href{https://dx.doi.org/10.1103/PhysRevB.100.035311}{Phys. Rev. B
  {\bf 100}, 035311}~(2019).

\bibitem{Thomas2022}
Philip Thomas, Leonardo Ruscio, Olivier Morin, and Gerhard Rempe.
\newblock ``Efficient generation of entangled multi-photon graph states from a
  single atom''.
\newblock \href{https://dx.doi.org/10.1038/s41586-022-04987-5}{Nature {\bf
  608}, 677--681}~(2022).

\bibitem{Delteil2016}
Aymeric Delteil, Zhe Sun, Wei-bo Gao, Emre Togan, Stefan Faelt, and Ata{\c{c}}
  Imamo{\u{g}}lu.
\newblock ``Generation of heralded entanglement between distant hole spins''.
\newblock \href{https://dx.doi.org/10.1038/nphys3605}{Nat. Phys. {\bf 12},
  218--223}~(2016).

\bibitem{Stockill2017}
R.~Stockill, M.~J. Stanley, L.~Huthmacher, E.~Clarke, M.~Hugues, A.~J. Miller,
  C.~Matthiesen, C.~Le~Gall, and M.~Atat\"ure.
\newblock ``Phase-tuned entangled state generation between distant spin
  qubits''.
\newblock \href{https://dx.doi.org/10.1103/PhysRevLett.119.010503}{Phys. Rev.
  Lett. {\bf 119}, 010503}~(2017).

\bibitem{Appel2022}
Martin~Hayhurst Appel, Alexey Tiranov, Simon Pabst, Ming~Lai Chan, Christian
  Starup, Ying Wang, Leonardo Midolo, Konstantin Tiurev, Sven Scholz,
  Andreas~D. Wieck, Arne Ludwig, Anders~S\o{}ndberg S\o{}rensen, and Peter
  Lodahl.
\newblock ``Entangling a hole spin with a time-bin photon: A waveguide approach
  for quantum dot sources of multiphoton entanglement''.
\newblock \href{https://dx.doi.org/10.1103/PhysRevLett.128.233602}{Phys. Rev.
  Lett. {\bf 128}, 233602}~(2022).

\bibitem{Browne2005}
Daniel~E. Browne and Terry Rudolph.
\newblock ``Resource-efficient linear optical quantum computation''.
\newblock \href{https://dx.doi.org/10.1103/PhysRevLett.95.010501}{Phys. Rev.
  Lett. {\bf 95}, 010501}~(2005).

\bibitem{Warburton2013}
Richard~J Warburton.
\newblock ``Single spins in self-assembled quantum dots''.
\newblock \href{https://dx.doi.org/10.1038/nmat3585}{Nat. Mater. {\bf 12},
  483--493}~(2013).

\bibitem{Lodahl2015}
Peter Lodahl, Sahand Mahmoodian, and S\o{}ren Stobbe.
\newblock ``Interfacing single photons and single quantum dots with photonic
  nanostructures''.
\newblock \href{https://dx.doi.org/10.1103/RevModPhys.87.347}{Rev. Mod. Phys.
  {\bf 87}, 347--400}~(2015).

\bibitem{Bernien2013}
Hannes Bernien, Bas Hensen, Wolfgang Pfaff, Gerwin Koolstra, Machiel~S Blok,
  Lucio Robledo, Tim~H Taminiau, Matthew Markham, Daniel~J Twitchen, Lilian
  Childress, et~al.
\newblock ``Heralded entanglement between solid-state qubits separated by three
  metres''.
\newblock \href{https://dx.doi.org/10.1038/nature12016}{Nature {\bf 497},
  86--90}~(2013).

\bibitem{Morley2017}
Sam Morley-Short, Sara Bartolucci, Mercedes Gimeno-Segovia, Pete Shadbolt, Hugo
  Cable, and Terry Rudolph.
\newblock ``Physical-depth architectural requirements for generating universal
  photonic cluster states''.
\newblock \href{https://dx.doi.org/10.1088/2058-9565/aa913b}{Quantum Sci.
  Technol. {\bf 3}, 015005}~(2017).

\bibitem{Zaporski2022}
Leon Zaporski, Noah Shofer, Jonathan~H Bodey, Santanu Manna, George Gillard,
  Martin~Hayhurst Appel, Christian Schimpf, Saimon~Filipe Covre~da Silva, John
  Jarman, Geoffroy Delamare, et~al.
\newblock ``Ideal refocusing of an optically active spin qubit under strong
  hyperfine interactions''.
\newblock \href{https://dx.doi.org/10.1038/s41565-022-01282-2}{Nat.
  Nanotechnol. {\bf 18}, 257–263}~(2023).

\bibitem{Nguyen2023}
Giang~N. Nguyen, Clemens Spinnler, Mark~R. Hogg, Liang Zhai, Alisa Javadi,
  Carolin~A. Schrader, Marcel Erbe, Marcus Wyss, Julian Ritzmann, Hans-Georg
  Babin, Andreas~D. Wieck, Arne Ludwig, and Richard~J. Warburton.
\newblock ``Enhanced electron-spin coherence in a gaas quantum emitter''.
\newblock \href{https://dx.doi.org/10.1103/PhysRevLett.131.210805}{Phys. Rev.
  Lett. {\bf 131}, 210805}~(2023).

\bibitem{Xu2007}
Xiaodong Xu, Yanwen Wu, Bo~Sun, Qiong Huang, Jun Cheng, D.~G. Steel, A.~S.
  Bracker, D.~Gammon, C.~Emary, and L.~J. Sham.
\newblock ``Fast spin state initialization in a singly charged inas-gaas
  quantum dot by optical cooling''.
\newblock \href{https://dx.doi.org/10.1103/PhysRevLett.99.097401}{Phys. Rev.
  Lett. {\bf 99}, 097401}~(2007).

\bibitem{Antoniadis2022}
Nadia~O Antoniadis, Mark~R Hogg, Willy~F Stehl, Alisa Javadi, Natasha Tomm,
  R{\"u}diger Schott, Sascha~R Valentin, Andreas~D Wieck, Arne Ludwig, and
  Richard~J Warburton.
\newblock ``Cavity-enhanced single-shot readout of a quantum dot spin within 3
  nanoseconds''.
\newblock \href{https://dx.doi.org/10.1038/s41467-023-39568-1}{Nat. Commun.
  {\bf 14}, 3977}~(2023).

\bibitem{Press2008}
David Press, Thaddeus~D Ladd, Bingyang Zhang, and Yoshihisa Yamamoto.
\newblock ``Complete quantum control of a single quantum dot spin using
  ultrafast optical pulses''.
\newblock \href{https://dx.doi.org/10.1038/nature07530}{Nature {\bf 456},
  218--221}~(2008).

\bibitem{Barrett2005}
Sean~D. Barrett and Pieter Kok.
\newblock ``Efficient high-fidelity quantum computation using matter qubits and
  linear optics''.
\newblock \href{https://dx.doi.org/10.1103/PhysRevA.71.060310}{Phys. Rev. A
  {\bf 71}, 060310(R)}~(2005).

\bibitem{Lim2005}
Yuan~Liang Lim, Almut Beige, and Leong~Chuan Kwek.
\newblock ``Repeat-until-success linear optics distributed quantum computing''.
\newblock \href{https://dx.doi.org/10.1103/PhysRevLett.95.030505}{Phys. Rev.
  Lett. {\bf 95}, 030505}~(2005).

\bibitem{Duan2005}
L.-M. Duan and R.~Raussendorf.
\newblock ``Efficient quantum computation with probabilistic quantum gates''.
\newblock \href{https://dx.doi.org/10.1103/PhysRevLett.95.080503}{Phys. Rev.
  Lett. {\bf 95}, 080503}~(2005).

\bibitem{Choi2019}
Hyeongrak Choi, Mihir Pant, Saikat Guha, and Dirk Englund.
\newblock ``Percolation-based architecture for cluster state creation using
  photon-mediated entanglement between atomic memories''.
\newblock \href{https://dx.doi.org/10.1038/s41534-019-0215-2}{npj Quantum
  Information {\bf 5}, 104}~(2019).

\bibitem{Denning2019}
Emil~V. Denning, Dorian~A. Gangloff, Mete Atat\"ure, Jesper M\o{}rk, and Claire
  Le~Gall.
\newblock ``Collective quantum memory activated by a driven central spin''.
\newblock \href{https://dx.doi.org/10.1103/PhysRevLett.123.140502}{Phys. Rev.
  Lett. {\bf 123}, 140502}~(2019).

\bibitem{Pompili2021}
Matteo Pompili, Sophie~LN Hermans, Simon Baier, Hans~KC Beukers, Peter~C
  Humphreys, Raymond~N Schouten, Raymond~FL Vermeulen, Marijn~J Tiggelman,
  Laura dos Santos~Martins, Bas Dirkse, et~al.
\newblock ``Realization of a multinode quantum network of remote solid-state
  qubits''.
\newblock \href{https://dx.doi.org/10.1126/science.abg1919}{Science {\bf 372},
  259--264}~(2021).

\bibitem{Gimeno2016}
Mercedes Gimeno-Segovia.
\newblock ``Towards practical linear optical quantum computing''.
\newblock PhD thesis.
\newblock Imperial College London.
\newblock ~(2016).
\newblock  url:~\href{https://doi.org/10.25560/43936}{doi.org/10.25560/43936}.

\bibitem{Herr2018}
Daniel Herr, Alexandru Paler, Simon~J Devitt, and Franco Nori.
\newblock ``A local and scalable lattice renormalization method for ballistic
  quantum computation''.
\newblock \href{https://dx.doi.org/10.1038/s41534-018-0076-0}{npj Quantum
  Information {\bf 4}, 27}~(2018).

\bibitem{Sykes1964}
M.~F. Sykes and John~W. Essam.
\newblock ``Exact critical percolation probabilities for site and bond problems
  in two dimensions''.
\newblock \href{https://dx.doi.org/10.1063/1.1704215}{Journal of Mathematical
  Physics {\bf 5}, 1117--1127}~(1964).

\bibitem{Hein2004}
M.~Hein, J.~Eisert, and H.~J. Briegel.
\newblock ``Multiparty entanglement in graph states''.
\newblock \href{https://dx.doi.org/10.1103/PhysRevA.69.062311}{Phys. Rev. A
  {\bf 69}, 062311}~(2004).

\bibitem{Hein2006}
Marc Hein, Wolfgang D{\"u}r, Jens Eisert, Robert Raussendorf, M~Nest, and H-J
  Briegel.
\newblock ``Entanglement in graph states and its applications''~(2006).
\newblock
  url:~\href{https://doi.org/10.48550/arXiv.quant-ph/0602096}{doi.org/10.48550/arXiv.quant-ph/0602096}.

\bibitem{VanderMarck1998}
Steven~C Van~der Marck.
\newblock ``Calculation of percolation thresholds in high dimensions for fcc,
  bcc and diamond lattices''.
\newblock \href{https://dx.doi.org/10.1142/S0129183198000431}{Int J Mod Phys C
  {\bf 9}, 529--540}~(1998).

\bibitem{Kurzawski2012}
{\L}ukasz Kurzawski and Krzysztof Malarz.
\newblock ``Simple cubic random-site percolation thresholds for complex
  neighbourhoods''.
\newblock \href{https://dx.doi.org/10.1016/S0034-4877(12)60036-6}{Rep. Math.
  Phys. {\bf 70}, 163--169}~(2012).

\bibitem{Lobl2023b}
Matthias~C. L{\"o}bl, Stefano Paesani, and Anders~S. S{\o}rensen.
\newblock ``Efficient algorithms for simulating percolation in photonic fusion
  networks''~(2023).
\newblock
  url:~\href{https://doi.org/10.48550/arXiv.2312.04639}{doi.org/10.48550/arXiv.2312.04639}.

\bibitem{Malarz2005}
Krzysztof Malarz and Serge Galam.
\newblock ``Square-lattice site percolation at increasing ranges of neighbor
  bonds''.
\newblock \href{https://dx.doi.org/10.1103/PhysRevE.71.016125}{Phys. Rev. E
  {\bf 71}, 016125}~(2005).

\bibitem{Xun2020b}
Zhipeng Xun and Robert~M. Ziff.
\newblock ``Bond percolation on simple cubic lattices with extended
  neighborhoods''.
\newblock \href{https://dx.doi.org/10.1103/PhysRevE.102.012102}{Phys. Rev. E
  {\bf 102}, 012102}~(2020).

\bibitem{Paesani2022}
Stefano Paesani and Benjamin~J. Brown.
\newblock ``High-threshold quantum computing by fusing one-dimensional cluster
  states''.
\newblock \href{https://dx.doi.org/10.1103/PhysRevLett.131.120603}{Phys. Rev.
  Lett. {\bf 131}, 120603}~(2023).

\bibitem{Newman2020}
Michael Newman, Leonardo~Andreta de~Castro, and Kenneth~R Brown.
\newblock ``Generating fault-tolerant cluster states from crystal structures''.
\newblock \href{https://dx.doi.org/10.22331/q-2020-07-13-295}{Quantum {\bf 4},
  295}~(2020).

\bibitem{Kramer1989}
Peter Kramer and Martin Schlottmann.
\newblock ``Dualisation of voronoi domains and klotz construction: a general
  method for the generation of proper space fillings''.
\newblock \href{https://dx.doi.org/10.1088/0305-4470/22/23/004}{Journal of
  Physics A: Mathematical and General {\bf 22}, L1097}~(1989).

\bibitem{Bell2022}
Thomas~J. Bell, Love~A. Pettersson, and Stefano Paesani.
\newblock ``Optimizing graph codes for measurement-based loss tolerance''.
\newblock \href{https://dx.doi.org/10.1103/PRXQuantum.4.020328}{PRX Quantum
  {\bf 4}, 020328}~(2023).

\bibitem{Economou2010}
Sophia~E. Economou, Netanel Lindner, and Terry Rudolph.
\newblock ``Optically generated 2-dimensional photonic cluster state from
  coupled quantum dots''.
\newblock \href{https://dx.doi.org/10.1103/PhysRevLett.105.093601}{Phys. Rev.
  Lett. {\bf 105}, 093601}~(2010).

\bibitem{Michaels2021}
Cathryn~P Michaels, Jes{\'u}s~Arjona Mart{\'\i}nez, Romain Debroux, Ryan~A
  Parker, Alexander~M Stramma, Luca~I Huber, Carola~M Purser, Mete Atat{\"u}re,
  and Dorian~A Gangloff.
\newblock ``Multidimensional cluster states using a single spin-photon
  interface coupled strongly to an intrinsic nuclear register''.
\newblock \href{https://dx.doi.org/10.22331/q-2021-10-19-565}{Quantum {\bf 5},
  565}~(2021).

\bibitem{Li2022}
Bikun Li, Sophia~E Economou, and Edwin Barnes.
\newblock ``Photonic resource state generation from a minimal number of quantum
  emitters''.
\newblock \href{https://dx.doi.org/10.1038/s41534-022-00522-6}{Npj Quantum Inf.
  {\bf 8}, 11}~(2022).

\bibitem{Stace2009}
Thomas~M. Stace, Sean~D. Barrett, and Andrew~C. Doherty.
\newblock ``Thresholds for topological codes in the presence of loss''.
\newblock \href{https://dx.doi.org/10.1103/PhysRevLett.102.200501}{Phys. Rev.
  Lett. {\bf 102}, 200501}~(2009).

\bibitem{Auger2018}
James~M. Auger, Hussain Anwar, Mercedes Gimeno-Segovia, Thomas~M. Stace, and
  Dan~E. Browne.
\newblock ``Fault-tolerant quantum computation with nondeterministic entangling
  gates''.
\newblock \href{https://dx.doi.org/10.1103/PhysRevA.97.030301}{Phys. Rev. A
  {\bf 97}, 030301(R)}~(2018).

\bibitem{Hastings2014}
Matthew~B. Hastings, Grant~H. Watson, and Roger~G. Melko.
\newblock ``Self-correcting quantum memories beyond the percolation
  threshold''.
\newblock \href{https://dx.doi.org/10.1103/PhysRevLett.112.070501}{Phys. Rev.
  Lett. {\bf 112}, 070501}~(2014).

\bibitem{Terhal2015}
Barbara~M. Terhal.
\newblock ``Quantum error correction for quantum memories''.
\newblock \href{https://dx.doi.org/10.1103/RevModPhys.87.307}{Rev. Mod. Phys.
  {\bf 87}, 307--346}~(2015).

\bibitem{Breuckmann2016}
Nikolas~P Breuckmann, Kasper Duivenvoorden, Dominik Michels, and Barbara~M
  Terhal.
\newblock ``Local decoders for the 2d and 4d toric code''~(2016).
\newblock
  url:~\href{https://doi.org/10.48550/arXiv.1609.00510}{doi.org/10.48550/arXiv.1609.00510}.

\bibitem{Breuckmann2021}
Nikolas~P. Breuckmann and Jens~Niklas Eberhardt.
\newblock ``Quantum low-density parity-check codes''.
\newblock \href{https://dx.doi.org/10.1103/PRXQuantum.2.040101}{PRX Quantum
  {\bf 2}, 040101}~(2021).

\bibitem{Tiurev2021b}
Konstantin Tiurev, Martin~Hayhurst Appel, Pol~Llopart Mirambell, Mikkel~Bloch
  Lauritzen, Alexey Tiranov, Peter Lodahl, and Anders~S\o{}ndberg S\o{}rensen.
\newblock ``High-fidelity multiphoton-entangled cluster state with solid-state
  quantum emitters in photonic nanostructures''.
\newblock \href{https://dx.doi.org/10.1103/PhysRevA.105.L030601}{Phys. Rev. A
  {\bf 105}, L030601}~(2022).

\bibitem{Nest2004}
Maarten Van~den Nest, Jeroen Dehaene, and Bart De~Moor.
\newblock ``Graphical description of the action of local clifford
  transformations on graph states''.
\newblock \href{https://dx.doi.org/10.1103/PhysRevA.69.022316}{Phys. Rev. A
  {\bf 69}, 022316}~(2004).

\bibitem{Looi2008}
Shiang~Yong Looi, Li~Yu, Vlad Gheorghiu, and Robert~B. Griffiths.
\newblock ``Quantum-error-correcting codes using qudit graph states''.
\newblock \href{https://dx.doi.org/10.1103/PhysRevA.78.042303}{Phys. Rev. A
  {\bf 78}, 042303}~(2008).

\bibitem{Zaidi2015}
Hussain~A. Zaidi, Chris Dawson, Peter van Loock, and Terry Rudolph.
\newblock ``Near-deterministic creation of universal cluster states with
  probabilistic bell measurements and three-qubit resource states''.
\newblock \href{https://dx.doi.org/10.1103/PhysRevA.91.042301}{Phys. Rev. A
  {\bf 91}, 042301}~(2015).

\bibitem{Cabello2011}
Ad\'an Cabello, Lars~Eirik Danielsen, Antonio~J. L\'opez-Tarrida, and Jos\'e~R.
  Portillo.
\newblock ``Optimal preparation of graph states''.
\newblock \href{https://dx.doi.org/10.1103/PhysRevA.83.042314}{Phys. Rev. A
  {\bf 83}, 042314}~(2011).

\bibitem{Adcock2020}
Jeremy~C Adcock, Sam Morley-Short, Axel Dahlberg, and Joshua~W Silverstone.
\newblock ``Mapping graph state orbits under local complementation''.
\newblock \href{https://dx.doi.org/10.22331/q-2020-08-07-305}{Quantum {\bf 4},
  305}~(2020).

\bibitem{Kok2010}
Pieter Kok and Brendon~W. Lovett.
\newblock ``Introduction to optical quantum information processing''.
\newblock \href{https://dx.doi.org/10.1017/CBO9781139193658}{Cambridge
  university press}. ~(2010).

\bibitem{Aaronson2004}
Scott Aaronson and Daniel Gottesman.
\newblock ``Improved simulation of stabilizer circuits''.
\newblock \href{https://dx.doi.org/10.1103/PhysRevA.70.052328}{Phys. Rev. A
  {\bf 70}, 052328}~(2004).

\bibitem{Fowler2009}
Austin~G. Fowler, Ashley~M. Stephens, and Peter Groszkowski.
\newblock ``High-threshold universal quantum computation on the surface code''.
\newblock \href{https://dx.doi.org/10.1103/PhysRevA.80.052312}{Phys. Rev. A
  {\bf 80}, 052312}~(2009).

\bibitem{Gottesman1998}
Daniel Gottesman.
\newblock ``Theory of fault-tolerant quantum computation''.
\newblock \href{https://dx.doi.org/10.1103/PhysRevA.57.127}{Phys. Rev. A {\bf
  57}, 127--137}~(1998).

\bibitem{Lobl2023}
Matthias~C. L{\"o}bl et~al.
\newblock ``perqolate''.
\newblock \url{https://github.com/nbi-hyq/perqolate}~(2023).

\bibitem{Conway1997}
John~H. Conway and Neil J.~A. Sloane.
\newblock ``Low--dimensional lattices. vii. coordination sequences''.
\newblock \href{https://dx.doi.org/10.1098/rspa.1997.0126}{Proceedings of the
  Royal Society of London. Series A: Mathematical, Physical and Engineering
  Sciences {\bf 453}, 2369--2389}~(1997).

\bibitem{Malarz2022_b}
Krzysztof Malarz.
\newblock ``Percolation thresholds on a triangular lattice for neighborhoods
  containing sites up to the fifth coordination zone''.
\newblock \href{https://dx.doi.org/10.1103/PhysRevE.103.052107}{Phys. Rev. E
  {\bf 103}, 052107}~(2021).

\bibitem{Malarz2022}
Krzysztof Malarz.
\newblock ``Random site percolation on honeycomb lattices with complex
  neighborhoods''.
\newblock \href{https://dx.doi.org/10.1063/5.0099066}{Chaos: An
  Interdisciplinary Journal of Nonlinear Science {\bf 32}, 083123}~(2022).

\bibitem{Derrida1985}
B.~Derrida and D.~Stauffer.
\newblock ``Corrections to scaling and phenomenological renormalization for
  2-dimensional percolation and lattice animal problems''.
\newblock \href{https://dx.doi.org/10.1051/jphys:0198500460100162300}{Journal
  de Physique {\bf 46}, 1623--1630}~(1985).

\bibitem{Mertens2018}
Stephan Mertens and Cristopher Moore.
\newblock ``Percolation thresholds and fisher exponents in hypercubic
  lattices''.
\newblock \href{https://dx.doi.org/10.1103/PhysRevE.98.022120}{Phys. Rev. E
  {\bf 98}, 022120}~(2018).

\bibitem{Feng2008}
Xiaomei Feng, Youjin Deng, and Henk W.~J. Bl{\"o}te.
\newblock ``Percolation transitions in two dimensions''.
\newblock \href{https://dx.doi.org/10.1103/PhysRevE.78.031136}{Phys. Rev. E
  {\bf 78}, 031136}~(2008).

\bibitem{Xu2014}
Xiao Xu, Junfeng Wang, Jian-Ping Lv, and Youjin Deng.
\newblock ``Simultaneous analysis of three-dimensional percolation models''.
\newblock \href{https://dx.doi.org/10.1007/s11467-013-0403-z}{Frontiers of
  Physics {\bf 9}, 113--119}~(2014).

\bibitem{Lorenz1998}
Christian~D. Lorenz and Robert~M. Ziff.
\newblock ``Precise determination of the bond percolation thresholds and
  finite-size scaling corrections for the sc, fcc, and bcc lattices''.
\newblock \href{https://dx.doi.org/10.1103/PhysRevE.57.230}{Phys. Rev. E {\bf
  57}, 230--236}~(1998).

\bibitem{Xun2020}
Zhipeng Xun and Robert~M. Ziff.
\newblock ``Precise bond percolation thresholds on several four-dimensional
  lattices''.
\newblock \href{https://dx.doi.org/10.1103/PhysRevResearch.2.013067}{Phys. Rev.
  Res. {\bf 2}, 013067}~(2020).

\bibitem{Hu2021}
Yi~Hu and Patrick Charbonneau.
\newblock ``Percolation thresholds on high-dimensional ${D}_{n}$ and
  ${E}_{8}$-related lattices''.
\newblock \href{https://dx.doi.org/10.1103/PhysRevE.103.062115}{Phys. Rev. E
  {\bf 103}, 062115}~(2021).

\bibitem{Morley2019}
Sam Morley-Short, Mercedes Gimeno-Segovia, Terry Rudolph, and Hugo Cable.
\newblock ``Loss-tolerant teleportation on large stabilizer states''.
\newblock \href{https://dx.doi.org/10.1088/2058-9565/aaf6c4}{Quantum Science
  and Technology {\bf 4}, 025014}~(2019).

\end{thebibliography}

\makeatletter 
\renewcommand{\thefigure}{A\@arabic\c@figure}
\makeatother

\makeatletter 
\renewcommand{\thetable}{A\@arabic\c@table}
\makeatother

\makeatletter 
\renewcommand{\theequation}{A\@arabic\c@equation}
\makeatother

\section{Appendix}

\subsection{Generating the resource states}
\label{appendix:resource}
\begin{figure*}
\includegraphics[width=1.0\textwidth]{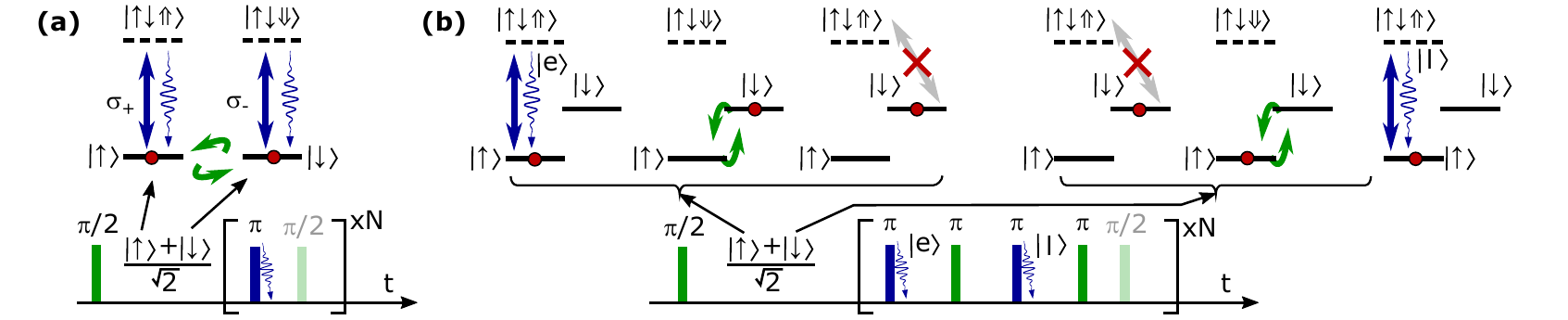}
\caption{\label{fig:resource_generate}\textbf{(a)} Pulse sequence to generate a linear cluster state with polarization encoding~\cite{Lindner2009}. By leaving out the second $\pi/2$-pulse (light green) one creates a GHZ state. When starting with spin $\ket{\uparrow}$, a photon with $\sigma_+$ polarization is emitted. When starting with spin $\ket{\downarrow}$, a photon with $\sigma_-$ polarization is emitted. \textbf{(b)} Corresponding pulse sequence for generating a linear cluster state or GHZ state with time-bin encoding~\cite{Tiurev2021}. When starting in the state $\ket{\uparrow}$, a photon $\ket{e}$ is emitted in the early time-bin. When starting in the state $\ket{\downarrow}$, no photon is generated in the early time-bin, yet in the late time-bin $\ket{l}$ after the spin has been flipped.}
\end{figure*}
As we have pointed out in the main text quantum emitters with a spin degree of freedom can be used to generate GHZ states or linear cluster states. This concept has been the subject of various theoretical proposals~\cite{Gheri1998,Lindner2009,Buterakos2017,Tiurev2021b,Tiurev2021} and some experimental realizations have recently been achieved~\cite{Schwartz2016,Coste2022,Cogan2023,Thomas2022}. In this work, star-shaped resource states with a central spin qubit are used and we therefore explain how such states can be created. The typical idea is to make use of a level scheme where an optical $\pi$-pulse on the emitter in the $\ket{\uparrow}$ state leads to a photon in state $\ket{0}$ which is orthogonal to the photon state $\ket{1}$ that is generated when starting with the spin state $\ket{\downarrow}$. The typical protocol uses polarization encoding~\cite{Lindner2009,Schwartz2016,Thomas2022,Coste2022,Cogan2023} (see Fig.~\ref{fig:resource_generate}(a)), but other degrees of freedom such as time-bin encoding can be used as well~\cite{Tiurev2021,Tiurev2021b} (see Fig.~\ref{fig:resource_generate}(b)). To generate a resource state, the spin is initialized in the state $\ket{+}_s=1/\sqrt{2}(\ket{\uparrow}+\ket{\downarrow})_s$. Applying $N$ $\pi$-pulses therefore leads to the following GHZ state:
\begin{equation}
\label{Eq:GHZ}
\frac{1}{\sqrt{2}}\left(\ket{\uparrow}_s\otimes\ket{0}^{\otimes N}+\ket{\downarrow}_s\otimes\ket{1}^{\otimes N}\right)
\end{equation}
Applying a single-qubit Hadamard gate $H$ to all photonic qubits transforms this state into:
\begin{align}
&\frac{1}{\sqrt{2}}\left(\ket{\uparrow}_s\otimes\ket{+}^{\otimes N}+\ket{\downarrow}_s\otimes\ket{-}^{\otimes N}\right) \\ & = \prod_{j=1}^N C_{sj}\left(\ket{+}_s\otimes\ket{+}^{\otimes N}\right),
\label{Eq:GHZ_after_H}
\end{align}
with $\ket{\pm}=1/\sqrt{2}(\ket{0}\pm\ket{1})$ and $C_{sj}$ representing a controlled-$Z$ gate between the spin and photon number $j$. This state is exactly a star-shaped graph state with the spin as the central qubit.

Graph states and GHZ states are stabilizer states~\cite{Nest2004,Hein2004}, and creating a star-shaped resource state from a GHZ state can thus be explained by transforming stabilizers. The stabilizer generators~\footnote{The stabilizers $\mathcal{S}$ of the stabilizer state $\ket{G}$ are all operators from the Pauli-group such that $\forall_{S\in\mathcal{S}}S\ket{G}=+1\ket{G}$. The stabilizer generators $\mathcal{S}_g$ are a subset of $N$ (number of qubits) independent elements of $\mathcal{S}$ that generate $\mathcal{S}$ ($\text{rank}(\mathcal{S})=\text{rank}(\mathcal{S}_g)=N$)~\cite{Hein2004}.} of the GHZ state \eqref{Eq:GHZ} are:
\begin{equation}
\braket{X\otimes X^{\otimes N}, Z\otimes Z_j},
\end{equation}
where the first operator acts on the spin qubit and the index $j\in\{1..N\}$ labels the corresponding photon. Applying $H^{\otimes N}$ to the GHZ state transforms its stabilizers in the Heisenberg picture. This leads to new stabilizer generators of the form
\begin{equation}
\label{Eq:stabilizer}
\braket{X\otimes Z^{\otimes N}, Z\otimes X_j},
\end{equation}
which are the stabilizers of the star-shaped resource state with a central spin qubit and $N$ photonic leaves~\footnote{That these operators are the stabilizers generators follows from the X-Z-rule~\cite{Looi2008}: Applying the operator $X$ to one qubit in a graph state is identical to applying the operator $Z$ to all its neighbors. Thus, applying both operations simultaneously acts as identity on the graph state as $X\cdot X=Z\cdot Z =1$}.

Note that the state in Eq.\ \eqref{Eq:GHZ} can be transferred to a purely photonic GHZ state by measuring the spin in the $X$-basis. Measuring $\ket{+}_s$ projects the system into the corresponding $N$-photon GHZ state. When measuring $\ket{-}_s$, an additional $Z$-gate needs to be applied to an arbitrary photon to retain the same photonic GHZ state. Therefore, photonic star-shaped resource states can be generated with the above method as well.

\subsection{Fusion circuits}
\label{appendix:fusion}
\begin{figure}[t!]
\includegraphics[width=1.0\columnwidth]{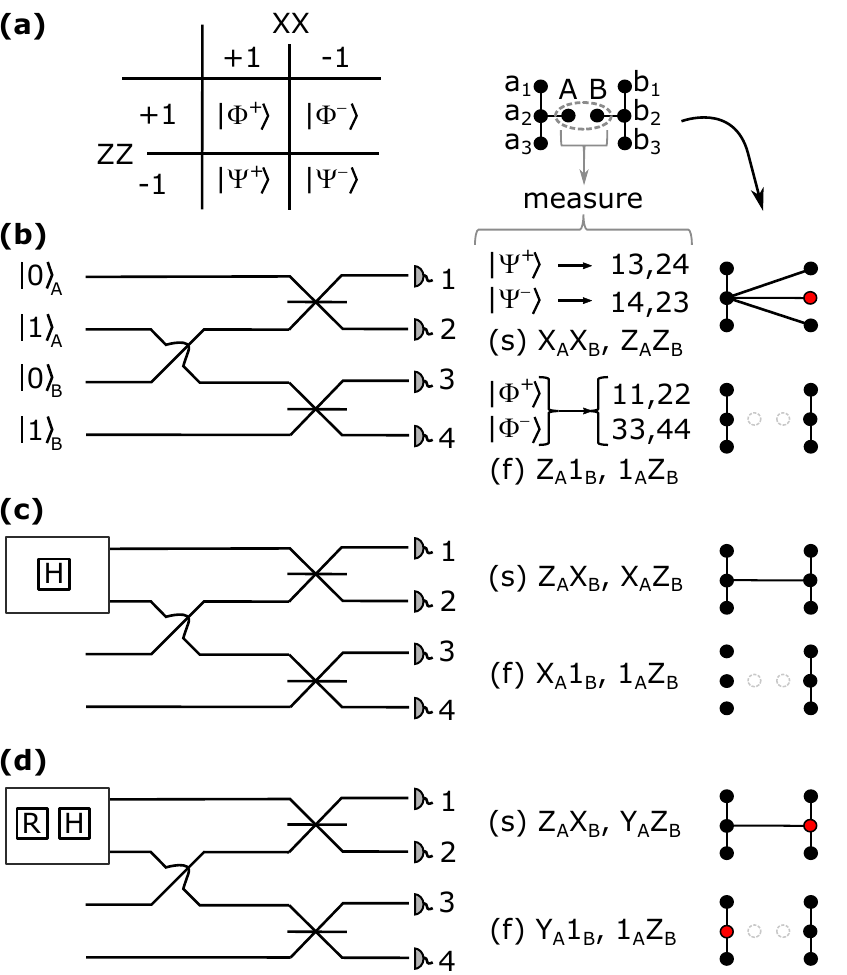}
\caption{\label{fig:fusion}\textbf{(a)} The four Bell basis states are stabilized by the operators $\pm XX$ and $\pm ZZ$. We consider the effect of fusing the qubits A and B in the graph that is shown on the right. We consider three different circuits in the following. \textbf{(b)} Upon success (s), the depicted circuit (left half) measures the Bell states $\ket{\Psi^{\pm}}$ and therefore the corresponding stabilizers. The Bell states $\ket{\Phi^{\pm}}$ result in the same detection patterns and therefore cannot be distinguished (fusion failure). In the failure case (f), the circuit measures both qubits in the $Z$-basis. Fusion success and failure transform the graph state as illustrated on the right. To obtain the displayed graph state, the gate $H$ (Hadamard) needs to be applied after the fusion on the qubit highlighted in red. \textbf{(c)} Rotated fusion with the gate $H$ applied to the first qubit before the fusion. \textbf{(d)} Modified fusion that can be used to merge star-shaped resource states into cluster states. $R=\big(\begin{smallmatrix}1 & 0\\0 & i\end{smallmatrix}\big)$ is applied post-fusion to one qubit (red) to obtain the shown graph state.}
\end{figure}

In this section, we describe the fusion operations that we assume for fusing star-shaped resource states (such as the two states in Fig.~\ref{fig:fusion}(a)). The process is known as type-II fusion~\cite{Browne2005} which, upon success, corresponds to two simultaneous parity measurements involving the detection of two photons. For pedagogical reasons we first describe two fusion operations that may be useful for other applications~\cite{Zaidi2015}, yet do not correspond to the fusion model in Figs.~\ref{fig:scheme}(b,c). We then will give a detailed explanation of a rotated Bell measurement that has exactly the desired properties~\cite{GimenoSegovia2015}.

The fusion circuit in Fig.~\ref{fig:fusion}(b) is a Bell measurement where the states $\ket{\Psi^{\pm}}=\frac{1}{\sqrt{2}}(\ket{0_A1_B}\pm\ket{1_A0_B})$ are measured upon a specific detection pattern. Sending, for instance, the state $\ket{\Psi^{+}}$ to the depicted fusion circuit results in the detection patterns 13 or 24, which are only measured for this particular Bell basis state (here e.g. 13 corresponds to clicks in detectors 1 and 3). $\ket{\Psi^{+}}$ and $\ket{\Psi^{-}}$ are eigenstates of $-Z_AZ_B$ and $\pm X_AX_B$ with eigenvalues $+1$ (and these operators are known as the stabilizers of the states). A successful fusion operation corresponds to a projection on $\ket{\Psi^{+}}$ or $\ket{\Psi^{-}}$ and thus is a simultaneous measurement of the stabilizers $-Z_AZ_B$ and $\pm X_AX_B$. The states $\ket{\Phi^{\pm}}=\frac{1}{\sqrt{2}}(\ket{0_A0_B}\pm\ket{1_A1_B})$ cannot be distinguished by the circuit as they both result in the same detection patterns 11, 22, 33, or 44\footnote{Note that we assume photon-number resolving detectors here such that the detection pattern 11, for instance, can be distinguished from a single photon in mode 1 and a lost photon elsewhere.}. Their detection pattern corresponds to an independent measurement of both qubits in the Z-basis. Failure of the fusion operation thus corresponds to the measurement of the operators $Z_A\mathds{1}_B$ and $\mathds{1}_AZ_B$.

Simultaneously measuring $Z_AZ_B$ and $X_AX_B$ connects graph components that were detached before. However, it does not connect star-shaped resource states in the desired way, since it produces the graph state illustrated in Fig.~\ref{fig:fusion}(b), which is not locally equivalent to the desired graph in Fig.~\ref{fig:fusion}(d)~\cite{Cabello2011,Adcock2020}. (Note that the derivation for the corresponding graph transformation is very similar to the derivation of the rotated fusion that we will discuss in the course of this section.)

The effect of the fusion circuit can be modified by adding single qubit gates before the Bell measurement~\cite{Zaidi2015,GimenoSegovia2015}. When applying such a gate $U$, the stabilizers $S_i$ of the graph state get rotated in the Heisenberg picture: $S_i\rightarrow US_iU^{\dag}$. To understand the effect on the graph state, one can analyze the effect of the Bell measurement on the rotated stabilizers. Alternatively one can rotate the Bell measurement as well as the stabilizers by applying $U^{\dag}$ from the left and $U$ from the right. This operation leaves the stabilizers the same ($U^{\dag}US_iU^{\dag}U=S_i$) and rotates the Bell measurement: when $U$ acts on qubit A, the measured operators $X_AX_B$ and $Z_AZ_B$ become $U^{\dag}X_AU\otimes X_B$ and $U^{\dag}Z_AU\otimes X_B$. Upon failure, the measured operators become $U^{\dag}Z_AU\otimes \mathds{1}_B$ and $\mathds{1}_AZ_B$

A typical modification is rotating the Bell measurement by adding a single Hadamard gate before one of the two qubits~\cite{Kok2010,Zaidi2015}. Upon fusion success, the operators $X_AZ_B$ and $Z_AX_B$ are then measured which corresponds to the desired fusion operation on the graph state~\cite{Kok2010,Zaidi2015} (see Fig.~\ref{fig:fusion}(c)). Upon failure, however, it corresponds to measuring qubit $A$ in the $X$- and qubit $B$ in the $Z$-basis. The $X$-measurement of one fusion qubit also measures the central qubit of the corresponding star-shaped state in the $Z$-basis\footnote{The reason is that the resource state in Fig.~\ref{fig:fusion}(a) before the fusion has a stabilizer of the form $X_AZ_{a_2}$. Measuring $X_A$ therefore automatically measures $Z_{a_2}$ as the product $X_AZ_{a_2}$ must be equal to $+1$.}~\cite{Hein2006}. This $Z$-measurement destroys one star-shaped resource state as illustrated in Fig.~\ref{fig:fusion}(c). For this reason, the corresponding fusion circuit is not suited for our purposes.

A suitable fusion circuit has been given in Refs.~\cite{GimenoSegovia2015,Gimeno2016} for polarization-encoded qubits. We give a detailed explanation of a corresponding circuit for dual-rail encoding which is based on transforming stabilizers upon the measurements~\cite{Aaronson2004,Fowler2009} (fusions). The circuit is shown in Fig.~\ref{fig:fusion}(d) where the gate $R=\sqrt{Z}=\big(\begin{smallmatrix}1 & 0\\0 & i\end{smallmatrix}\big)$ followed by a Hadamard gate $H=\frac{X+Z}{\sqrt{2}}$ is applied to one of the fusion qubits~\footnote{$R$ interchanges Pauli $X$, $Y$ and leaves $Z$ the same: $RXR^{\dag}=Y$, $RYR^{\dag}=-X$, $RZR^{\dag}=Z$~\cite{Gottesman1998}. $H$ interchanges Pauli $X$, $Z$ and leaves $Y$ the same: $HXH^{\dag}=Z$, $HYH^{\dag}=-Y$, $HZH^{\dag}=X$~\cite{Gottesman1998}. The circuit in Fig.~\ref{fig:fusion}(d) applies $K^{\dag}:=H\cdot R$ to fusion qubit $A$. This gate cyclically interchanges the Pauli matrices: $K^{\dag}XK=-Y$, $K^{\dag}YK=-Z$, $K^{\dag}ZK=X$ ($KXK^{\dag}=Z$, $KYK^{\dag}=-X$, $KZK^{\dag}=-Y$). Different gate choices are possible. Ref.~\cite{GimenoSegovia2015} uses $R:=\frac{X-Y}{\sqrt{2}}=i\cdot\exp\left(-i\frac{\pi}{2}\cdot(n_xX + n_yY + n_zZ)\right)$ with $\Vec{n}=\frac{1}{\sqrt{2}}\left(\Vec{e}_x-\Vec{e}_y\right)$ which transforms the Pauli operators in the same way (up to signs): $RXR^{\dag}=-Y$, $RYR^{\dag}=-X$, $RZR^{\dag}=-Z$. Here, we can neglect the signs as they do not affect the structure of the stabilizers (resp. transform the graph state to a graph basis state~\cite{Looi2008}). The signs can be compensated post-fusion in measurement-based quantum computing or by applying a Pauli $Z$-gate to all qubits $i$ for which the stabilizer $S_i$ has a minus sign ($ZXZ^{\dag}=-X$).}. Applying this gate $K^{\dag}:=H\cdot R$ to one of the fusion qubits has the effect of transforming the Bell measurement from $X_AX_B$, $Z_AZ_B$ to $Z_AX_B$, $Y_AZ_B$ upon success and from $Z_A\mathds{1}_B$, $\mathds{1}_AZ_B$ to $Y_A\mathds{1}_B$, $\mathds{1}_AZ_B$ upon failure.

Before the fusion, the graph state depicted in Fig.~\ref{fig:fusion}(a) has the following stabilizers:
\begin{align}
&S_{a_1} = X_{a_1}Z_{a_2}\\
&S_{a_2} = X_{a_2}Z_{a_1}Z_{a_3}Z_A \label{stab_ZA}\\
&S_{a_3} = X_{a_3}Z_{a_2}\\
&S_{A} = X_AZ_{a_2}\label{stab_XA}\\
&S_{b_1} = X_{b_1}Z_{b_2}\\
&S_{b_2} = X_{b_2}Z_{b_1}Z_{b_3}Z_B\\
&S_{b_3} = X_{b_3}Z_{b_2}\\
&S_{B} = X_BZ_{b_2}
\end{align}
Upon fusion success, the circuit in Fig.~\ref{fig:fusion}(d) measures $Z_AX_B$ and $Y_AZ_B$ so only stabilizers commuting with these operators remain invariant. This obviously applies to all stabilizers with no support on qubits $A$, $B$. From the four remaining stabilizers, two that commute with both measured operators can be obtained by multiplying $S_{a_2}$ with $S_{B}$ as well as $S_A$ with $S_{B}$ and $S_{b_2}$. These multiplications yield the following remaining stabilizers: 
\begin{align}
&X_{a_2}Z_{a_1}Z_{a_3}Z_{b_2}Z_AX_B\\
&Z_{a_2}Y_{b_2}Z_{b_1}Z_{b3}X_AY_B\\
&X_{a_1}Z_{a_2}\\
&X_{a_3}Z_{a_2}\\
&X_{b_1}Z_{b_2}\\
&X_{b_3}Z_{b_2}
\end{align}
Omitting the Pauli matrices on the destructively measured photonic qubits $A$ and $B$ yields the following stabilizers:
\begin{align}
&\Tilde{S}_{a_2} = X_{a_2}Z_{a_1}Z_{a_3}Z_{b_2}\\
&R_{b_2}\Tilde{S}_{b_2}R^{\dag}_{b_2} = Z_{a_2}Y_{b_2}Z_{b_1}Z_{b3}\\
&\Tilde{S}_{a_1} = X_{a_1}Z_{a_2}\\
&\Tilde{S}_{a_3} = X_{a_3}Z_{a_2}\\
&\Tilde{S}_{b_1} = X_{b_1}Z_{b_2}\\
&\Tilde{S}_{b_3} = X_{b_3}Z_{b_2}
\end{align}
Applying the gate $R_{b_2}$ after the Bell-measurement changes $Y_{b_2}$ into $X_{b_2}$ and leaves the rest identical. The resulting operators $\title{S}_i$ are the stabilizer generators of the graph state in Fig.~\ref{fig:fusion}(d).

Upon fusion failure, the circuit measures the operators $Y_A\mathds{1}_B$ and $\mathds{1}_AZ_B$. Multiplying the stabilizer $X_{a_2}Z_{a_1}Z_{a_3}Z_A$ from Eq.~\ref{stab_ZA} with $X_AZ_{a_2}$ from Eq.~\ref{stab_XA} yields $Y_{a_2}Z_{a_1}Z_{a_3}Y_A$ which is commuting with the measurement pattern of the fusion failure. The stabilizers commuting with the measurement pattern are (omitting the measured fusion qubits):
\begin{align}
&\Tilde{S}_{a_1} = X_{a_1}Z_{a_2}\\
&\Tilde{S}_{a_3} = X_{a_3}Z_{a_2}\\
&\Tilde{S}_{b_1} = X_{b_1}Z_{b_2}\\
&\Tilde{S}_{b_3} = X_{b_3}Z_{b_2}\\
&R_{a_2}\Tilde{S}_{a_2}R^{\dag}_{a_2} = Y_{a_2}Z_{a_1}Z_{a_3}\\
&\Tilde{S}_{b_2} = X_{b_2}Z_{b_1}Z_{b_3}
\end{align}
When applying the gate $R_{a_2}$ after the fusion, these stabilizers correspond to the graph state for the failure mode of Fig.~\ref{fig:fusion}(d).

Note that the central qubits $a_2$ and $b_2$ might be affected by previous fusions. However, the above calculation remains unaffected when the initial stabilizers (before the fusion) contain the operators $Y_{a_2}$ or $Y_{b_2}$ from previous fusions (instead of $X_{a_2}$ or $X_{b_2}$). When another fusion involving the leaf qubits $a_1$, $a_3$, $b_1$, or $b_3$ is performed, then $Y_{a_2}$ or $Y_{b_2}$ may change back to $X_{a_2}$ resp. $X_{b_2}$. To obtain a graph state, only the qubits that have a stabilizer containing a Pauli $Y$ rather than a Pauli $X$ matrix need a gate $R$ after the fusions. As all these gates can be performed after all fusions are done, the need to apply $R$-gates (resp. changing the measurement basis correspondingly for the subsequent measurement-based quantum computing~\cite{Raussendorf2001}) does not compromise the concept of applying all fusion operations at once (ballistically).

\subsection{Definition of lattices}
\label{appendix:lattice_def}
\begin{figure*}
\includegraphics[width=1.0\textwidth]{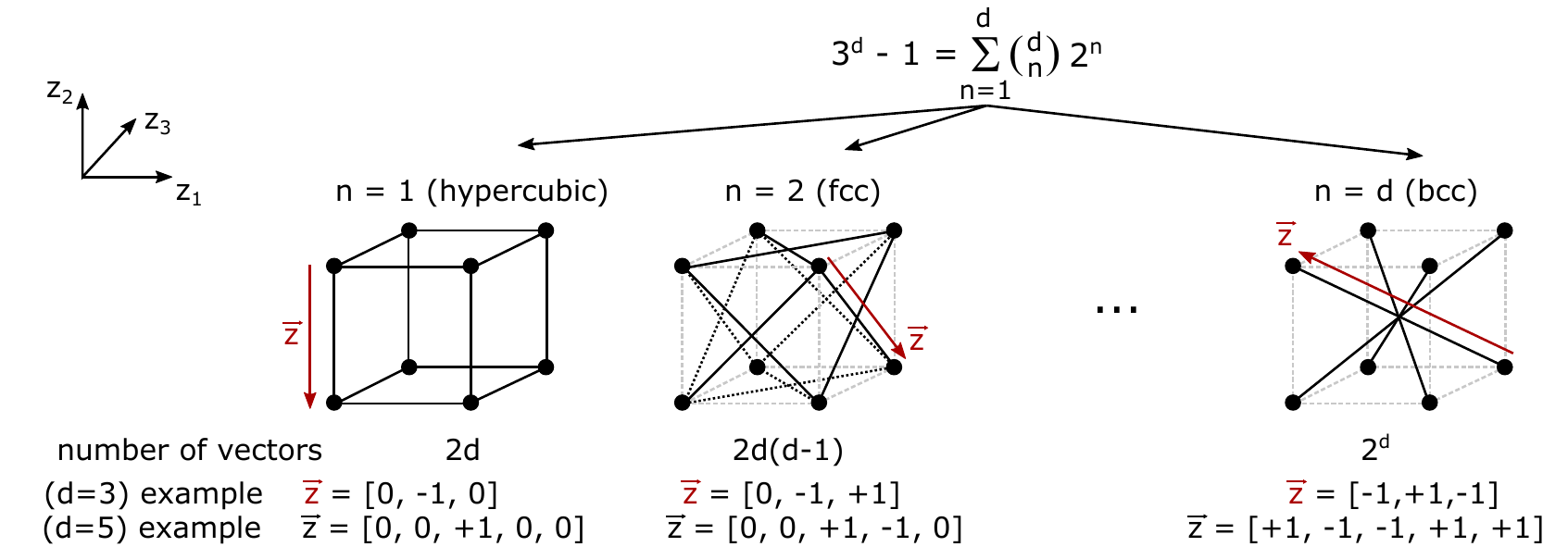}
\caption{\label{fig:lattice}Construction of different lattices by adding connection vectors to the lattice $\mathbb{Z}^d$. The illustration shows three-dimensional lattices, but the construction works in any dimension. In the illustration, every connection vector $\vec{z}$ fulfills $|z_i|\leq k=1$ for all its elements, resp. $\vec{z}\in\{0,\pm1\}^d$. At the same time, the number of dimensions in which the connection vector is non-zero is fixed to a specific value $n=\sum_{i=1}^d|z_i|$. With this definition, the hypercubic lattice corresponds to $n=1$, the fcc lattice corresponds to $n=2$, and the bcc lattice corresponds to $n=d$. For all these cases, an exemplary connection vector $\vec{z}$ is given in red.}
\end{figure*}
In the main text, we have studied the loss tolerance of several lattices on which star-shaped resource states are geometrically arranged and fused. In this section, we describe how the corresponding lattices are mathematically defined. Our source code implementing these lattices is openly available at Ref.~\cite{Lobl2023}.

Graphs with a periodic lattice structure can be created by locally connecting vertices and there are several techniques to create various classes of lattices~\cite{Conway1997,VanderMarck1998,Kurzawski2012}. In the main text, we have explained a lattice construction that starts with arranging unconnected vertices on the hypercubic lattice $\mathbb{Z}^d$. We then specify a set of integer vectors associated with the connections between the vertices. For most lattices that we consider, all nodes/vertices are equivalent (so a connection vector $\vec{z}\in\mathbb{Z}^d$ has a corresponding partner $-\vec{z}$ in the opposite direction). When the maximum integer value per dimension is restricted by $|z_i|\leq k$, all connection vectors are in the set $\Set{\vec{z}\in\mathbb{Z}^d: \abs{z_i}\leq k} \setminus \Set{0}$ and thus there are in total $(2k+1)^{d}-1$ different connection vectors.

Even when just one step per dimension is allowed ($k=1$), various lattices can be constructed by combining connection vectors, as illustrated in Fig.~\ref{fig:lattice}: the hypercubic lattice is made by all $2d$ connection vectors with a non-zero element in only one dimension ($\sum_{i=1}^d|z_i|=1$); the fcc lattice is made by all $2d(d-1)$ connection vectors with a non-zero element in exactly two dimensions ($\sum_{i=1}^d|z_i|=2$); and the bcc lattice is made by all $2^{d}$ connection vectors with a non-zero step in every dimension ($\sum_{i=1}^d|z_i|=d$). Intuitively speaking, the hypercubic lattice corresponds to every node being connected to its first nearest neighbor, while the fcc lattice corresponds to every node being connected to its second nearest neighbor.

With this method, we construct all lattices from Ref.~\cite{VanderMarck1998} except the Kagome lattice as well as lattices with complex neighborhoods (see e.g. Refs.~\cite{Kurzawski2012,Malarz2022_b,Malarz2022}). In our simulations, we consider the hypercubic, the bcc, and the fcc lattice as well as combinations of them~\footnote{The $d$-dimensional bcc lattice alone consists of $2^{d-1}$ mutually unconnected partitions, and the fcc lattice alone has $d(d-1)$ mutually unconnected components. Combining these lattices with the hypercubic lattice makes them fully connected.}. In three dimensions, for instance, the combination of hypercubic (simple cubic) and fcc lattice leads to the lattice NN+2NN~\cite{Kurzawski2012}.

Except for the diamond lattice, all the lattices that we consider can be constructed with the above method. A graph with the structure of the generalized diamond lattice can be created by removing edges from the hypercubic lattice~\cite{VanderMarck1998}. In this construction, all hypercubic edges pointing in the first dimension ($|z_1|=1$) stay untouched. Every second edge with a connection vector stepping in any of the other dimensions is removed. In particular, when $\vec{z}$ connects the vertices $V^{A}$ and $V^{B}$ with $V^{A}$ having the smaller coordinate in the dimension $l$ ($V^{A}_l < V^{B}_l$, $|z_l|=1$, $l>1$), the corresponding edge $e=(V^{A}, V^{B})$ is removed iff $\text{mod}(V^{A}_l,2)=0$.

\subsection{Classical percolation simulations}
\label{appendix:classical}
\begin{table*}
\begin{centering}
\small
\begin{tabular}{|c c c c c c|}
 \hline
 \textbf{$\lambda_c^{site}$}& d=2 & d=3 & d=4 & d=5 & d=6 \\
 \hline\hline
 diamond & 0.6968(7) & 0.4302(19) & 0.2968(27) & 0.2197(39) & 0.1733(14)\\
 & 0.6971(2)\cite{VanderMarck1998} & 0.4301(2)\cite{VanderMarck1998} & 0.2978(2)\cite{VanderMarck1998} & 0.2252(3)\cite{VanderMarck1998} & 0.1799(5)\cite{VanderMarck1998}\\
 \hline
 hc & 0.5929(3)& 0.3119(4) & 0.1973(10) & 0.1402(5) & 0.1079(13)\\
 & 0.59274(10)\cite{Derrida1985} & 0.3116(1)\cite{Kurzawski2012} & 0.19688561(3)\cite{Mertens2018} & 0.14079633(4)\cite{Mertens2018} & 0.109016661(8)\cite{Mertens2018}\\
\hline
 fcc & 0.5928(8)& 0.1995(3)& 0.0842(2)& 0.0430(2)& 0.0252(6)\\
 & 0.59274(10)\cite{Derrida1985} & 0.1994(2)\cite{VanderMarck1998,Kurzawski2012} & 0.0842(3)\cite{VanderMarck1998} & 0.0431(3)\cite{VanderMarck1998} & 0.0252(5)\cite{VanderMarck1998}\\
\hline
 bcc & 0.5928(9) & 0.2460(6)& 0.1028(13)& 0.0426(25)& 0.0184(5)\\
 & 0.59274(10)\cite{Derrida1985} & 0.2458(2)\cite{VanderMarck1998,Kurzawski2012} & 0.1037(3)\cite{VanderMarck1998} & 0.0446(4)\cite{VanderMarck1998} & 0.0199(5)\cite{VanderMarck1998}\\
 \hline
 fcc+hc & 0.4069(10) & 0.1376(1) & 0.0616(5) & 0.0334(2) & 0.0207(2)\\
  & 0.40725395(3)\cite{Feng2008} & 0.1372(1)\cite{Kurzawski2012} & & & \\
\hline
 bcc+hc & 0.4074(9) & 0.1360(5) & 0.0587(6) & 0.0290(1) & 0.0147(4)\\
 & 0.40725395(3)\cite{Feng2008}  & $\sim$0.136~\footnote{Reviewed site-percolation threshold for the NN+3NN lattice, private communication, K. Malarz, 2023}, 0.1420(1)~\cite{Kurzawski2012} & & & \\
 \hline
 \hline
 \textbf{$\lambda_c^{bond}$} & d=2 & d=3 & d=4 & d=5 & d=6 \\
 \hline\hline
 diamond & 0.6529(10) & 0.3893(7) & 0.2709(18) & 0.2072(15) & 0.1646(43) \\
 & 0.652703645...\cite{Sykes1964} & 0.3893(2)\cite{VanderMarck1998,Xu2014} & 0.2715(3)\cite{VanderMarck1998} & 0.2084(4)\cite{VanderMarck1998} & 0.1677(7)\cite{VanderMarck1998} \\
 \hline
 hc & 0.5000(5) & 0.2491(2) & 0.1602(7) & 0.1185(6) & 0.0940(8)\\
 & 1/2 & 0.2488126(5)\cite{Lorenz1998} & 0.1601312(2)\cite{Xun2020b} & 0.11817145(3)\cite{Mertens2018} & 0.09420165(2)\cite{Mertens2018} \\
\hline
 fcc & 0.4996(4) & 0.1202(1) & 0.0495(2) & 0.0271(3) & \\
 & 1/2 & 0.1201635(10)\cite{Lorenz1998} & 0.049517(1)\cite{Xun2020} & 0.0271813(2)\cite{Hu2021} & \\
\hline
 bcc & 0.5003(7) & 0.1801(2) & 0.0741(1) & 0.0320(12) & 0.0142(2)\\
 & 1/2 & 0.1802875(10)\cite{Lorenz1998} & 0.074212(1)\cite{Xun2020} & 0.033(1)\cite{VanderMarck1998} & \\
 \hline
 fcc+hc & 0.2505(5) & 0.0753(2) & 0.0359(1) & 0.0213(1) & \\
  & 0.25036834(6)\cite{Feng2008} & 0.0752326(6)\cite{Xun2020b} & 0.035827(1)\cite{Xun2020} & & \\
\hline
 bcc+hc & 0.2505(10) & 0.0921(2) & 0.0457(3) & 0.0244(3) & 0.0122(5)\\
 & 0.25036834(6)\cite{Feng2008} & 0.0920213(7)\cite{Xun2020b} & & & \\
 \hline
\end{tabular}
\caption{Results of classical site-percolation ($\lambda_c^{site}$, upper part) and bond-percolation ($\lambda_c^{bond}$, lower part) simulations of various lattices together with corresponding literature values. Note that some of the lattices are identical, for instance, hc and bcc are the same in two dimensions.}
\label{table:classical}
\end{centering}
\end{table*}
In our simulation, we first build a certain lattice, second, apply a particular percolation model, and third analyze the resulting graph structure. To verify that our implementation of various lattices in the first step is correct, we perform classical bond- and site-percolation simulations of these lattices and compare the resulting values to the literature~\cite{VanderMarck1998,Kurzawski2012}. The site-percolation thresholds, $\lambda_c^{site}$, resulting from these simulations are shown in the upper part of Table~\ref{table:classical}. The simulated lattices are mainly the lattices described in section~\ref{appendix:lattice_def}. Additionally, we simulate the lattices 2NN+3NN and NN+2NN+3NN~\cite{Kurzawski2012} (lattices with complex neighborhoods) for which we find $\lambda_c^{site}=0.1039(6)$ and $\lambda_c^{site}=0.0978(1)$, respectively (in reasonable agreement with the literature values $0.1036(1)$, $0.0976(1)$~\cite{Kurzawski2012}).

To compensate for finite size effects in our simulations, we use the approach from Ref.~\cite{VanderMarck1998} but more elaborate methods that do not rely on cluster spanning could be used instead~\cite{Malarz2022_b}. Most percolation thresholds agree very well with the literature values. The only exceptions are the six-dimensional bcc and diamond lattices. The deviation by a few standard deviations might be caused by an underestimation of the systematic error (by us and/or the literature~\cite{VanderMarck1998}) when using the method from Ref.~\cite{VanderMarck1998} to compensate for finite size effects.

The lower part of Tab.~\ref{table:classical} shows corresponding bond-percolation simulations. In the case of no photon loss, the resulting percolation thresholds, $\lambda_c^{bond}$, specify the fusion success probability that is minimally required for creating a large percolating cluster state~\cite{Paesani2022}. Also for the bond-percolation thresholds, we find a good agreement with the literature values.

\subsection{Percolation thresholds}
\label{appendix:thresh}
In Fig.~\ref{fig:2} of the main text, we plot percolation thresholds for an architecture where cluster states are built from star-shaped resource states. In Table~\ref{table:threshold} we give the corresponding numerical values for the case that the central qubit of the star-shaped resource states is either a spin or a photon. The percolation thresholds for the purely photonic architecture are higher compared to the case of a central spin qubit because of unheralded photon loss. When unheralded loss is present, the corresponding percolation thresholds present only a necessary condition for quantum computing after lattice renormalization, because unheralded loss cannot be foreseen during the renormalization/path-finding~\cite{Morley2019}.

\begin{table*}
\begin{centering}
\begin{tabular}{|c| c c c c c|}
 \hline
 lattice & d=2 & d=3 & d=4 & d=5 & d=6\\
 \hline
 diamond & $-$ & 0.9639(1) & 0.9314(4) & 0.9194(3) & 0.9145(11)\\
 \hline
 hc & 1.0 & 0.9435(1) & 0.9300(1) & 0.9281(3) & 0.9292(8)\\
 \hline
 fcc+hc & 0.9703(1) & 0.9574(1) & 0.9638(1) & 0.9691(2) & \\
\hline
 bcc+hc & 0.9703(1) & 0.9465(1) & 0.9522(1) & 0.9591(1) & \\
 \hline
 fcc & $-$ & 0.9513(1) & 0.9581(1) & 0.9647(8) & \\
 \hline
 bcc & $-$ & 0.9422(2) & 0.9454(1) & 0.9573(7) & \\
 \hline\hline
 lattice & d=2 & d=3 & d=4 & d=5 & d=6\\
 \hline
 diamond & $-$ & 0.9729(2) & 0.9475(3) & 0.9368(11) & 0.9329(6)\\
 \hline
 hc & 1.0 & 0.9561(1) & 0.9446(2) & 0.9428(3) & 0.9437(1)\\
 \hline
 fcc+hc & 0.9768(1) & 0.9653(1) & 0.9703(1) & 0.9746(5) & \\
\hline
 bcc+hc & 0.9768(1) & 0.9575(1) & 0.9616(3) & 0.9671(2) & \\
 \hline
 fcc & $-$ & 0.9605(1) & 0.9657(1) & 0.9710(3) & \\
 \hline
 bcc & $-$ & 0.9546(1) & 0.9563(2) & 0.9656(4) & \\
 \hline
\end{tabular}
\caption{Percolation thresholds $\lambda_c^{\eta}$ for the loss-tolerance of fusion lattices in several dimensions (with rotated type-II fusion and star-shaped resource states). In the \textbf{upper part}, the central qubit of every star-shaped resource state is a spin that cannot suffer unheralded loss. In the \textbf{lower part}, the star-shaped resource states are purely photonic and the central qubits can suffer unheralded loss. For the two-dimensional diamond brickwork lattice (honeycomb), no percolation is possible since the bond-percolation threshold of the lattice is above the fusion probability of 0.5~\cite{Browne2005}. The two-dimensional simple cubic (square) lattice has a bond percolation threshold of exactly 0.5 and $\eta=1.0$ is therefore required.}
\label{table:threshold}
\end{centering}
\end{table*}

\subsection{Size of the largest connected component}
\label{appendix:size}
\begin{figure*}[t]
\includegraphics[width=1.0\textwidth]{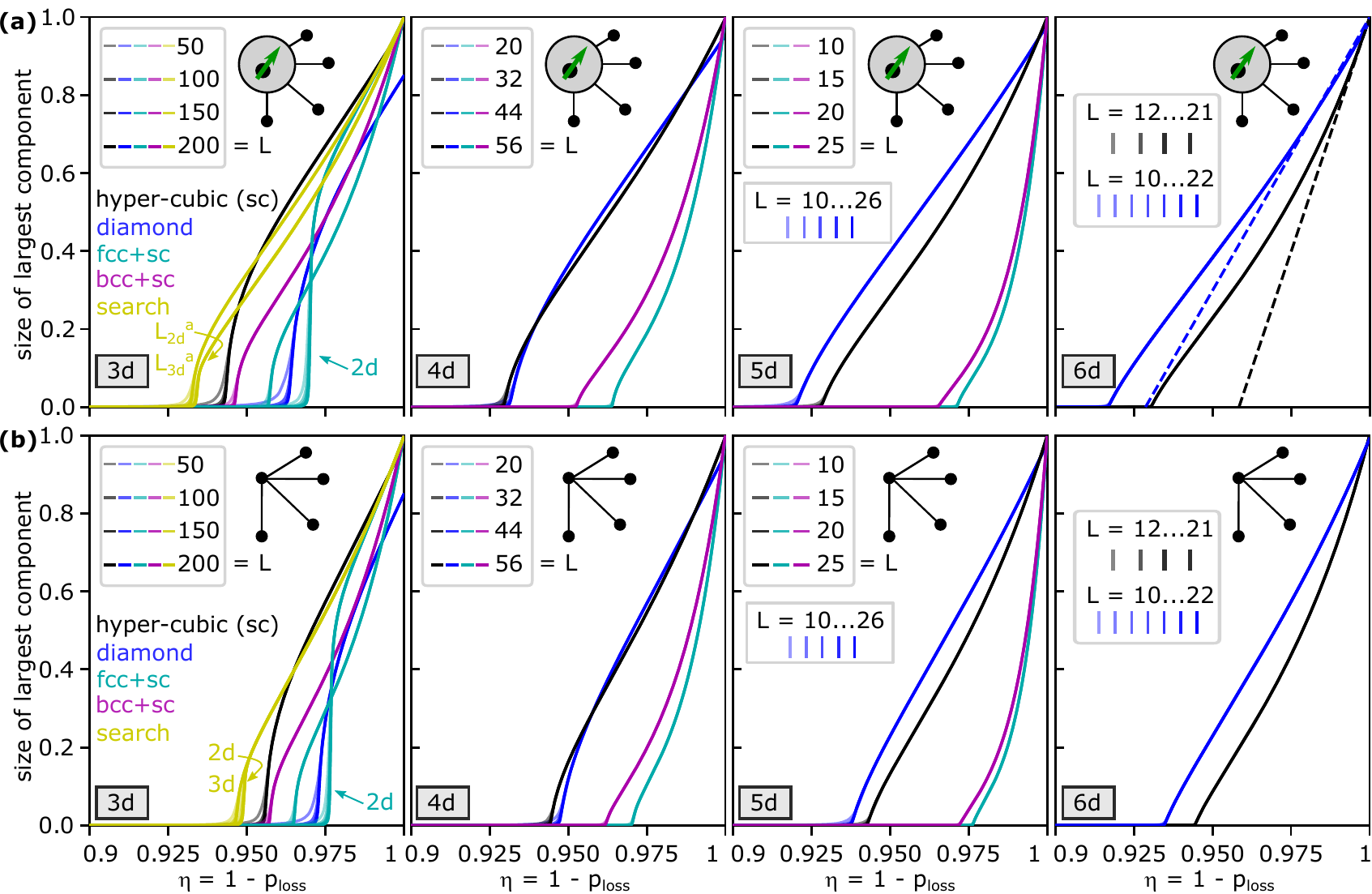}
\caption{\label{size_all}Size of the largest connected component relative to the initial number of star-shaped resource states in the fusion lattice. The largest component size is plotted against the efficiency $\eta=1-p_{loss}$. Smaller lattice sizes are indicated by a lighter color, but the curves almost overlap for the chosen lattice sizes. \textbf{(a)} Simulation with a spin qubit as the central qubit of every star-shaped resource state. The dashed lines illustrate how the number of active qubits in the six-dimensional hypercubic and diamond lattices decreases in the regime of low photon loss. \textbf{(b)} The corresponding simulations for purely photonic resource states.}
\end{figure*}
In the main text, we have presented percolation thresholds that specify the loss tolerance of different fusion lattices. When the efficiency $\eta=1-p_{loss}$ is above this threshold, the created graph state has a large connected component that grows linearly with the system size. From a practical point of view, it is interesting to know the size of the largest connected component of the graph. We simulate the largest connected component size as a function of $\eta$ using periodic boundary conditions (to reduce finite size effects). The results of these simulations are presented in Fig.~\ref{size_all}. When a precise number for $\eta$ is known in an experiment, such simulations show with which fusion lattice the largest connected graph state can be obtained. A smaller percolation threshold does not necessarily mean that the lattice performs better for all values of $\eta$. An example is the four-dimensional diamond lattice which has a slightly higher percolation threshold than the four-dimensional hypercubic lattice, yet performs better in terms of component size in a certain range of $\eta$.

An interesting point is that the size of the largest connected component decreases linearly for very low losses. The slope of this decrease is directly connected to the vertex degree $d_v$ of the lattice. Consider a fusion lattice made of star-shaped resource states with a central spin qubit in a quantum emitter. If $|V|$ is the number of spins, then the number of fusion photons is $d_v|V|$. On average, the number of lost photons is then $d_v\cdot|V|\cdot p_{loss}$. Every lost photon on a fusion edge $E$ will make both spins connected to $E$ useless (as they are measured in the Z-basis). For very low photon losses, there are practically no spin qubits that are affected by two photon losses simultaneously and the overall number of active spin qubits in the graph thus becomes $|V|\cdot(1 - 2\cdot d_v\cdot p_{loss})$. The slope with which the relative size of the largest connected component decreases at low losses is, therefore, $- 2\cdot d_v\cdot p_{loss}$. We have illustrated this slope for the six-dimensional hypercubic lattice and the six-dimensional diamond lattice in Fig.~\ref{size_all}(a). The six-dimensional hypercubic lattice has a vertex degree of $d_v=12$ and the corresponding slope matches the simulation well. For the diamond lattice, the match is less good which we attribute to its lower vertex degree of $d_v=8$ in combination with the finite fusion success probability. Fusion failure leads to missing bonds and the size of the largest connected component can therefore be smaller than the number of active spins (some spins may be unconnected even for zero loss). 

When the photon loss increases, the slope of the largest connected component as a function of loss can change for two reasons. (1) For more losses, the chance that the same spin qubit is affected by two or more different photon losses increases. Consequently, fewer spin qubits are affected and the slope bends towards less negative values (compared to $- 2\cdot d_v\cdot p_{loss}$) lowering the percolation threshold. This effect can be seen for lattices with high vertex degrees such as the five-dimensional fcc+hc lattice. The high vertex degree increases the chance that a qubit is affected by two losses. (2) When a lattice is already heavily affected by photon loss, both loss and the finite fusion probability may cause a large component to be split into two pieces. In this case, the size of the largest component rapidly decreases which causes the slope to become more negative (compared to $- 2\cdot d_v\cdot p_{loss}$) making the percolation threshold higher. This effect can be seen for the three-dimensional lattices in Fig.~\ref{size_all}.

\subsection{Results of lattice search}
\label{appendix:lattice}
\begin{table*}
\begin{centering}
\begin{tabular}{|c| c| c|}
 \hline
 lattice & vectors $\{\vec{z}^{(j)}\}$ representing lattice & $\lambda_c^{\eta}$\\
 \hline
 $L_{2d}^a$ & \{[5,7], [7,4], [0,3], [7,-6], [7,-5]\} & 0.9344(1)\\
 $L_{2d}^b$ & \{[1,2], [6,-6], [1,4], [4,-5]\} & 0.9362(3)\\
 $L_{2d}^c$ & \{[4,-3], [7,1], [4,4], [7,2]\} & 0.9352(1)\\
 \hline
 $L_{3d}^a$ & \{[1,0,1], [1,-1,-1], [1,1,0], [1,1,-1]\} & 0.9326(2)\\
 $L_{3d}^b$ & \{[1,2,-2], [2,0,1], [1,-1,-1], [2,-1,0]\} & 0.9307(5)\\
 \hline
 $hc_{4d}$ & \{[1,0,1,0], [1,0,0,1], [1,0,0,0], [1,-1,0,-1]\} & 0.9295(2)\\
 \hline\hline
 dimension & vectors $\{\vec{z}^{(j)}\}$ representing lattice & $\lambda_c^{\eta}$\\
 \hline
 d=2 & \{[1,1], [6,-7], [3,1], [7,-2]\} & 0.9493(2)\\
 \hline
 d=3 & \{[1,1,-1], [0,1,1], [1,-1,1], [1,-1,0]\} & 0.9468(1)\\
 \hline
 d=4 & \{[1,-1,0,1], [1,-1,-1,1], [1,1,0,-1], [1,-1,1,-1], [1,1,0,0]\} & 0.9444(7)\\
 \hline
\end{tabular}
\caption{Fusion lattices with particularly high tolerance to photon loss and the corresponding percolation thresholds. For the upper part of the table, the central qubit of each star-shaped resource state is a spin. For the lower part, the resource states are purely photonic.}
\label{table:lattice_search}
\end{centering}
\end{table*}
We have performed a search for particularly loss-tolerant fusion lattices with a greedy optimization described in the main text. Every lattice is represented by a set of connection vectors $\{\vec{z}^{(j)}\}$ that specify connections between lattice points $\mathbb{Z}^d$. Note that the presence of a connection vector $\vec{z}^{(j)}$ means that there is also a connection along the vector $-\vec{z}^{(j)}$ since we only consider lattices where all nodes are equivalent. So if we give say 4 different vectors $\vec{z}^{(j)}$, then $d_v=2\cdot4$.

The results of this search in dimensions two, three, and four are given in Table~\ref{table:lattice_search} together with the corresponding percolation thresholds. First, we consider the case that the central qubits of the resource states are spins. In two dimensions, we allow connection vectors with $|\vec{z}^{(j)}_i| \leq7$ and we find the lattices $L_{2d}^a$, $L_{2d}^b$, $L_{2d}^c$ (with $L_{2d}^a$ being shown in Fig.~\ref{fig:lattices}(b) of the main text). In three dimensions, we allow all connection vectors with $|\vec{z}^{(j)}_i| \leq1$ and we find the lattice $L_{3d}^a$ which is shown in Fig.~\ref{fig:lattices}(c) of the main text. Allowing $|\vec{z}^{(j)}_i| \leq2$ leads to the slightly better lattice $L_{3d}^{b}$. The optimized four-dimensional lattice (with $|\vec{z}^{(j)}_i| \leq1$) is equivalent to the hypercubic lattice up to distortions. Finally, we also consider purely photonic resource states. The results of these simulations are shown in the lower part of Table~\ref{table:lattice_search}.

\end{document}